\pdfoutput=1
\documentclass[aps,prfluids,fixfloat,onecolumn]{revtex4-2}
\usepackage{graphicx,amsmath,amsfonts,bm, color}
\usepackage{soul}
\usepackage{comment}

\newcommand*\colvec[3][]{\begin{pmatrix}\ifx\relax#1\relax\else#1\\\fi#2\\#3\end{pmatrix}}
\newcommand*\rowvec[3][]{\begin{pmatrix}\ifx\relax#1\relax\else#1&\fi#2&#3\end{pmatrix}}

\newcommand{\mycomment}[1]{}
\newcommand{\mcB}{\mathcal{B}}
\newcommand{\mcC}{\mathcal{C}}
\newcommand{\mcR}{\mathcal{R}}

\newcommand{\deleted}[1]{\iffalse{\color{red}{\st{#1}}}\fi}

\usepackage{hyperref}
\begin{document}

\title{Enucleated incompressible red blood cells in shear flow: \\ theoretical analysis of shape instabilities}

\author{Avraham Moriel$^1$, Howard A. Stone$^1$, and Simon Mendez$^2$}
\affiliation{$^1$ Department of Mechanical and Aerospace Engineering, Princeton University, Princeton, New Jersey 08544, USA \\ 
$^2$ Institut Montpelliérain Alexander Grothendieck, CNRS, University of Montpellier, Montpellier 34095, France}

\begin{abstract}
Red blood cells (RBCs) are essential for oxygen transport, and their remarkable ability to undergo significant deformations during flow is a crucial feature for their physiological function. At intermediate shear rates typical of the microcirculation, RBCs can adopt complex, multi-lobed shapes, signifying a dynamic instability. Here we adopt a perturbative theoretical framework of a quasi-spherical RBC under external shear flow to study such shape instabilities. To better capture RBC maturation and enucleation, we first extend the framework to explicitly account for different excess areas between the stress-free and current membrane shapes. We revisit the reduced equations of motion obtained for an ellipsoidally-shaped RBC, and demonstrate the effect of different excess areas and initial orientation on the dynamical trajectories. Then, we introduce additional spatial modes and show that an emerging instability critically depends on the RBC's shear and bending moduli, the internal to external viscosity ratio, and the excess area, mainly through the RBC's membrane tension. We also study the instability-induced saturation of  the membrane tension, and the resulting excess area redistribution at long times. The theoretical framework and the emerging picture of the different instabilities provide insights into the emergence of stomatocyte and trilobe shapes exhibited by RBCs under external flow.

\end{abstract}
\maketitle

\section{Introduction}\label{se:intro}

Red blood cells (RBCs) are the primary cellular constituents of blood and are fundamental to the maintenance of healthy circulation. Due to their abundance and deformability, RBCs play a central role in the rheological properties of blood, enabling efficient flow and oxygen transport through the body's vasculature~~\cite{chien1970shear,merrill1966blood,lanotte2016red}. RBCs mature within the bone marrow, undergoing a critical process of enucleation where the cell nucleus is expelled~\cite{keerthivasan2011erythroblast,lu2008biologic,menon2021erythroid}. Once released into the blood stream, RBCs typically adopt a biconcave discocyte shape and form a suspension of deformable particles~\cite{fischer2004shape}. This suspension allows proper transport of oxygen and nutrients
and is vital for physiological function~\cite{secomb1991red,goldsmith1972flow}. 

The dynamics of individual RBCs have long been a subject of study not only for their physiological relevance but also for their rich dynamics, serving as a model system for soft matter under flow~\cite{guest1963red,barthes2016motion,dupire2012full,goldsmith1972flow,kaoui2009red,dupire2010chaotic,abkarian2008vesicles,fischer1978red,fischer1978tank,abkarian2007swinging,dupire2015simple,mendez2018plane,jeffery1922motion,keller1982motion,skotheim2007red}. When subjected to shear flows, RBCs display a variety of behaviors: RBCs can exhibit rigid-body-like tumbling~\cite{dupire2012full,goldsmith1972flow}, where the entire cell rotates; tank treading, where the membrane continuously circulates around a fixed interior~\cite{fischer1978red,fischer1978tank}; or swinging, characterized by oscillations around a steady orientation~\cite{abkarian2007swinging}. These diverse dynamics motivated the development of several theoretical frameworks. Since RBCs largely maintain their overall biconcave discocyte shape during such simple motions, the models treated RBCs as rigid or nearly rigid objects~\cite{jeffery1922motion,keller1982motion,skotheim2007red}. These rigid-body theories, including those based on the classical Jeffery orbits, successfully predicted the transitions between tumbling, tank-treading, and swinging dynamics~\cite{abkarian2007swinging,abkarian2008vesicles,dupire2012full,dupire2015simple,mendez2018plane}. 

Recently, experiments reported a wealth of complex RBC shapes emerging at intermediate shear rates that cannot be explained by rigid-body models. Specifically, unusual morphologies such as the symmetry-broken stomatocytes and various multi-lobed shapes were observed under external shear rates, typically ranging from $100$ to $1000$ s$^{-1}$~\cite{fischer1978tank, lanotte2016red}. Although several numerical studies explored the deformation of RBCs under flow~\cite{mauer2018flow,abbasi2022dynamics}, the underlying physical mechanisms driving these shape changes remain debated. Specifically, as multi-lobed shapes were observed in both 3D RBC or 2D vesicle simulations, it is unclear what roles are played by elasticity, bending, membrane tension, excess area, and the internal and external viscosities, on initiating shape instabilities, and how they affect the emerging morphologies~\cite{fischer1978tank,lanotte2016red,mauer2018flow,abbasi2022dynamics}. 

Here, we theoretically explore the emergence of different RBC shapes under shear flow.
To this end, we revisit a theoretical framework that fully accounts for membrane mechanics and deformations, and describes the RBC’s spatio-temporal shape dynamics, enabling an analytical investigation of RBC stability in the quasi-spherical shape limit~\cite{vlahovska2011dynamics, vlahovska2007dynamics, olla2000behavior, misbah2006vacillating,danker2007dynamics,kaoui2009vesicles}: 
Vlahovska \emph{et. al}~\cite{vlahovska2011dynamics} presented the linearization of the full, non-linear framework using a perturbative expansion, derived the equations of motion for the leading-order components, analyzed a simplified case of an ellipsoidal RBC and demonstrated dynamics of different, well-known scenarios (e.g., capsules and vesicles).
By incorporating different excess areas for the stress-free and current shapes --- inspired by RBC's enucleation during maturation --- and by including additional spatial modes, we probe the emerging RBC dynamics, instabilities, and the emerging shape configurations at long times, for a wide range of material parameters. This allows us to study the interplay between membrane tension, elasticity, and bending resistance. As membrane tension is the only mechanism that couples the spatial modes at leading order in our framework, the present study examines the role of membrane tension as the key driver of RBC shape instabilities.

We begin by stating the governing equations, dimensionless numbers, and parameterizations of the problem in Sec.~\ref{se:math}. We revisit the equations of motion of the reduced system, considering only the ellipsoidal deformation modes with different stress-free and current excess areas, and demonstrate the resulting dynamics in Sec.~\ref{se:upd}.  In Sec.~\ref{se:inst}, we introduce additional spatial modes to develop a linear instability criterion, whose validity and resulting phase diagram are verified numerically. Then, in Sec.~\ref{se:ss}, we analyze the long-time behavior, providing an analytical approximation for the excess area distribution after an instability occurs. We conclude by discussing the emerging physical picture and its implications in Sec.~\ref{se:discussion}.

\section{Mathematical framework}\label{se:math}
We treat the RBC as a closed, fluid-filled membrane suspended in an externally driven ambient fluid~\cite{vlahovska2011dynamics, vlahovska2007dynamics, olla2000behavior, misbah2006vacillating,danker2007dynamics,kaoui2009vesicles}. Both internal and external fluids are assumed to be incompressible, 
\begin{equation} \label{eq:full_incomp}
\nabla \cdot \bm{v}^{\text{ex/in}} = 0 \ ,
\end{equation}
where $\bm{v}^{\text{ex/in}}$ denote the external and internal velocity fields respectively. 

The momentum balance equation is greatly simplified, as the RBC's characteristic size leads to a low Reynolds number, and we neglect inertial contributions. The fluids are assumed to be Newtonian, resulting in Stokes equations:
\begin{equation} \label{eq:full_mombal}
    \nabla \cdot \bm{T}^{\text{ex/in}} = -\nabla p^{\text{ex/in}} + \eta^{\text{ex/in}} \nabla^2 \bm{v}^{\text{ex/in}} = \bm{0} \ , 
\end{equation}
where $\bm{T}^{\text{ex/in}}$ are the stress tensors, $\eta^{\text{ex/in}}$ are the external and internal viscosities, and $p^{\text{ex/in}}$ are the external and internal pressures, respectively.

The presence of the membrane imposes two essential sets of boundary conditions at the interface. First, the fluid velocities  $\bm{v}^{\text{ex/in}}$ evaluated on the membrane boundary $\left.\bullet\right|_m$ are assumed to be continuous across the membrane, and match the membrane velocity $\bm{v}^{\text{m}}$. Additionally, the forces exerted by the fluids balance with the membrane force density $\bm{\tau}^\text{m}$. These two conditions are cast as
\begin{equation} 
\label{eq:membrane_cont}
\left.\bm{v}^{\text{ex}}\right|_{\text{m}} = \left.\bm{v}^{\text{in}}\right|_{\text{m}} = \bm{v}^{\text{m}} \ , \qquad \qquad \left.\bm{n}\cdot\left(\bm{T}^{\text{ex}} - \bm{T}^{\text{in}} \right)\right|_{\text{m}} = \bm{\tau}^{\text{m}} \ ,
\end{equation}
where $\bm{n}$ is the outward unit normal to the membrane.

To proceed, we need to specify the contributions to the RBC membrane force density $\bm{\tau}^{\text{m}}$. The RBC's membrane is a composite structure, consisting of a lipid bilayer connected to a spectrin-based cytoskeletal network~\cite{lim2002stomatocyte,peng2014erythrocyte}. Consequentially, the resulting membrane force density is decomposed into three components: elastic, bending, and membrane tension:
\begin{equation} \label{eq:force_decomps}
    \bm{\tau}^{\text{m}} = \bm{\tau}^{\mu} + \bm{\tau}^{\kappa} + \bm{\tau}^{s} \ .
\end{equation}
Here we omit the potential contributions that arise due to membrane viscosity (these have been shown to reduce deformations and cause damped oscillatory deformation of the RBC~\cite{yazdani2013influence}).

The elastic response of the cytoskeletal network is captured using the Skalak constitutive model~\cite{skalak1973strain}. Denoting by $\bm{x}$ the current configuration, $\bm{X}$ the stress-free configuration, and $\bm{d}\!=\!\bm{x}-\bm{X}$ the displacement, the elastic force density can be approximated for small deformations
as a two-dimensional equivalent to Hooke's 
law~\cite{barthes1981time,vlahovska2011dynamics,brenner2013interfacial}
\begin{equation}\label{eq:mem_elastic}
\bm{\tau}^{\mu}=2\left(K_A - \mu\right)\left(\nabla_{s}\cdot\bm{d}\right)H\bm{n}-\left(K_A - \mu\right)\nabla_{s}\nabla_{s}\cdot\bm{d}-\mu\nabla_{s}\cdot\left[\nabla_{s}\bm{d}\cdot\bm{I}_{s}+\bm{I}_{s}\cdot\left(\nabla_{s}\bm{d}\right)^{T}\right] \ .
\end{equation}
Here $K_A$ is the area modulus and $\mu$ is the planar shear modulus, $H\!=\!\tfrac{1}{2}\nabla_s\cdot\bm{n}$ is the mean curvature, $\bm{I}_s\!=\!\bm{I}-\bm{n}\bm{n}$ is a projection operator onto the membrane's surface, and $\nabla_s\!=\!\bm{I}_s\cdot \nabla$ is the surface gradient~\cite{barthes1981time,vlahovska2007dynamics}.

The resistance of the lipid bilayer to bending is obtained from the Helfrich energy~\cite{helfrich1973elastic, seifert1997configurations}, yielding the bending force density $\bm{\tau}^{\kappa}$ as
\begin{equation} \label{eq:mem_bending}
    \bm{\tau}^{\kappa}\!=\!-\kappa\left(4H^{3}-4KH+2\nabla_{s}^{2}H\right)\bm{n}
\end{equation}
where $\kappa$ is the bending modulus, and $K$ is the Gaussian curvature.

Finally, the membrane tension contribution is given by~\cite{vlahovska2011dynamics,seifert1999fluid}
\begin{equation}\label{eq:mem_stension}
    \bm{\tau}^{s}\!=\!2\Sigma H\bm{n} - \nabla_s\Sigma \ .
\end{equation}
Here the membrane tension $\Sigma$ acts as a Lagrange multiplier enforcing constant local membrane area, and is not a material parameter. This implies $\Sigma$ varies spatially and temporally, and is determined as part of the solution. 

\begin{figure}[t]
\includegraphics[width=0.65\textwidth]{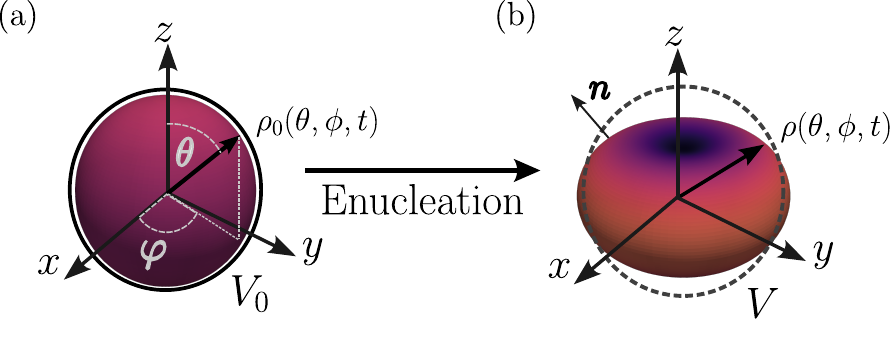}
\caption{A sketch of the enucleation of RBCs. (a) Before the enucleation the stress-free RBC takes an ellipsoidal shape with total surface area $A$, volume $V_0$, and an effective radius $R_0\!\equiv\!\left(3 V_0/4\pi\right)^{1/3}$, corresponding to excess area $\Delta_0$ (here $\Delta_0\!=\!0.01$). The stress-free surface is parametrized using $\rho_0(\theta,\varphi,t)$, where $\theta$ is the polar angle, and $\varphi$ is the azimuthal angle, and $t$ is the dimensionless time. (b) After enucleation the RBC acquires a new, discocyte shape with volume $V\!<\!V_0$ (depicted by the dashed circle), and an effective radius $R\!=\!\left(3 V / 4\pi\right)^{1/3}$, while the total surface area $A$ is conserved. This current shape is characterized by $\Delta\!$ (here $\Delta\!=\!1$), parametrized by $\rho(\theta,\varphi,t)$, with the outward unit normal to the membrane $\bm{n}$, and is axisymmetric around the $z$-axis. Colors heuristically indicate distance from the center of the stress-free and enucleated RBC respectively.}
\label{fig:fig1}
\end{figure}

A key aspect of our approach is the explicit consideration of the change in RBC geometry due to enucleation. Before the enucleation, the RBC is approximately an ellipsoid-shaped membrane of surface area $A$ enclosing a volume $V_0$. We define the typical stress-free radius $R_0$ of a sphere of a similar volume, i.e., $R_0\!\equiv\!\left(\frac{3 V_0}{4\pi}\right)^{1/3}$, and the initial excess area of the ellipsoid shape with respect to the sphere as $\Delta_0\!\equiv\!\frac{A}{R_0^2}-4\pi$. Then, the RBC undergoes enucleation, maturing into the discocyte shape. The resulting shape is a membrane of the same surface area $A$, but enclosing a volume $V\!<\!V_0$, with an effective radius $R\!=\!\left(\frac{3 V}{4\pi}\right)^{1/3}$, and excess area of $\Delta\!\equiv\!\frac{A}{R^2}-4\pi$; for typical RBC $\Delta\!\simeq\!4$~\cite{evans1972improved,levant2016intermediate,vlahovska2011dynamics}. As $V\!<\!V_0$, $R\!<\!R_0$, and $\Delta\!>\!\Delta_0$, the excess area is increased at the cost of locked internal stresses in the RBC membrane. The effect of reduced volume is schematically shown in Fig.~\ref{fig:fig1}. As we shall see, this increase in excess area alters the RBC response to external flow, and allows instabilities to emerge. 

Under an imposed shear flow $\bm{v}^\infty\!=\!\dot{\gamma} y \hat{\bm{x}}$ (here $\hat{\bm{x}}$ is a unit vector in the $x$-direction, and $\dot{\gamma}$ is the shear rate), the above description leads to several dimensionless numbers. The  internal-to-external viscosity ratio $\lambda\!\equiv\!\eta^{\text{in}}/\eta^{\text{ex}}$ plays a crucial role in RBC dynamics~\cite{levant2016intermediate,fischer2013threshold,abkarian2007swinging,secomb1991red,pfafferott1985red}. For a typical RBC of radius $R$, comparing the bending modulus $\kappa$, and the (2D) shear modulus $\mu$, to the typical stresses imposed by the external drive, we define the bending number $\mcB$ and the capillary number $\mcC$ as
\begin{equation} \label{eq:dimless_nums}
 \mcB\!=\!\frac{\eta^{\text{ex}} \dot{\gamma} R^3}{\kappa} \! \qquad \ , \qquad \mcC\!\equiv\!\frac{\eta^{\text{ex}} \dot{\gamma} R}{\mu} \ . 
\end{equation}
In what follows, we treat the RBC as an area-incompressible membrane, implying $\nabla_s\cdot\bm{d}\!=\!0$, rendering the area modulus $K_A$ irrelevant. 

To analytically explore the membrane dynamics, we employ a perturbative framework, expanding the RBC shape in spherical harmonics~\cite{vlahovska2011dynamics, vlahovska2007dynamics, olla2000behavior, misbah2006vacillating,danker2007dynamics,kaoui2009vesicles}. The current RBC membrane $\rho(\theta,\varphi,t)$, and the stress-free shape $\rho_0(\theta,\varphi,t)$, depicted in Fig.~\ref{fig:fig1}, are expanded in spherical harmonics $Y_{l,m}$'s, as~\cite{vlahovska2011dynamics}
\begin{subequations}\label{eq:rho_rho0}
    \begin{align}
     \rho(\theta,\varphi,t)\!=\! 1 + \sum_{l = 2}^{\infty} \sum_{m=-l}^l Y_{l,m} (\theta,\varphi) f_{l,m}(t) \ , \\
     \rho_0(\theta,\varphi,t) \!=\! 1 + \sum_{l = 2}^{\infty} \sum_{m=-l}^l  Y_{l,m}(\theta,\varphi) g_{l,m}(t) \ .
    \end{align}
\end{subequations}
Here we used $R$ and $R_0$, respectively, to rescale $\rho$ and $\rho_0$ into this dimensionless form. We assume that, while the RBC has undergone enucleation $\Delta_0\!<\!\Delta\!<\!4\pi$, the radii ratio is $\tfrac{R}{R_0}\!=\!\sqrt{\tfrac{\Delta_0+4\pi}{\Delta + 4\pi}}\simeq\!1$. The amplitudes $f_{l,m}(t)$ and $g_{l,m}(t)$ describe the current and stress-free shape dynamics (see Appendix~\ref{app:Ylmrot} for details about the spherical harmonics $Y_{l,m}$'s). The summation in Eq.~\eqref{eq:rho_rho0} starts at $l\!=\!2$ as we are interested in the RBC's deformations --- the $l\!=\!0$ modes correspond to volumetric modes (which are not allowed as the fluids are incompressible), and $l\!=\!1$ modes correspond to center of mass motion. 

The current and stress-free excess areas $\Delta$ and $\Delta_0$ are approximated as
\begin{equation} \label{eq:excess_f}
    \Delta \!=\! \sum_{l = 2}^{\infty} \sum_{m=-l}^l a_l f_{l,m}(t)f_{l,m}^*(t) + \mathcal{O}\left(f^3\right) \ , \qquad \Delta_0 \!=\! \sum_{l=2}^{\infty} \sum_{m=-l}^la_l g_{l,m}(t)g_{l,m}^*(t) + \mathcal{O}\left(g^3\right)\ , 
\end{equation}
where $a_l\!=\!(l+2)(l-1)/2$, $\bullet^*$ indicates the complex conjugate, and we emphasize that while $f_{l,m}(t)$ and $g_{l,m}(t)$ evolve in time, $\Delta$ and $\Delta_0$ remain constants. 
However, in the remainder of the paper, we suppressed the explicit time-dependence from  the $f_{l,m}$'s  and $g_{l,m}$'s for readability. These relations explicitly link the dynamic amplitudes $f_{l,m}$'s and $g_{l,m}$'s to their respective excess areas $\Delta$ and $\Delta_0$. Finally, the membrane tension is also expanded in a similar manner as 
\begin{equation} \label{eq:full_st}
\Sigma(\theta,\varphi,t)\!=\!\sigma(t) + \sum_{l=2}^{\infty} \sum_{m=-l}^l  Y_{l,m}\left(\theta,\varphi\right) \sigma_{l,m}(t) \ ,
\end{equation}
where $\sigma(t)$ is the spatially homogeneous part of the membrane tension and $\sigma_{l,m}(t)$'s are the time-dependent spatial projections of the membrane tension on the respective spherical harmonics $Y_{l,m}$'s.

The equations of motion are obtained by substituting the perturbative expansions of $\rho$, $\rho_0$, and $\Sigma$, together with the expressions for $\Delta$ and $\Delta_0$, Eqs.~\eqref{eq:rho_rho0}-\eqref{eq:full_st}, into the fluid dynamics Eqs.~\eqref{eq:full_incomp}-\eqref{eq:membrane_cont} and the membrane constitutive relations of $\bm{\tau}^\mu$, $\bm{\tau}^{\kappa}$, and $\bm{\tau}^s$, Eqs.~\eqref{eq:force_decomps}-\eqref{eq:mem_stension}, and collecting leading-order terms~\cite{vlahovska2011dynamics}. This results in a set of coupled equations for the spherical harmonic amplitudes $f_{l,m}$'s ~\cite{vlahovska2011dynamics}:

\begin{equation}\label{eq:eom}
\partial_{t}f_{l,m}\!=\!\left[i\omega m+\Upsilon _l(t)\right]f_{l,m}+S_{l,m}(t) \ ,
\end{equation}
where the equations are cast in dimensionless time $t$ (using the shear rate $\dot{\gamma}^{-1}$ as the time scale),  $\omega\!=\!\frac{1}{2}$ is the rotational frequency, and $\Upsilon_l(t)$ and $S_{l,m}(t)$ include the forces originating from both the fluid flow and membrane response as  
\begin{subequations} \label{eq:eom_sup}
\begin{align}
    \Upsilon_l(t) = & \mcB^{-1}\left[\Gamma_1^l + \sigma(t) \Gamma_2^l\right] + \mcC^{-1} \Gamma_3^l \ , \label{eq:eom_sup_ups}\\ 
    S_{l,m}(t) = & C_{l,m} - \mcC^{-1} \Gamma_3^l g_{l,m} \ .
\end{align}
\end{subequations}
Here $\Gamma_{1,2,3}^l$ are rational functions solely dependent on $l$ and $\lambda$, $\sigma(t)$ is the spatially homogeneous part of the membrane tension $\Sigma$ [see Eq.~\eqref{eq:full_st}], and $C_{l,m}$'s arise from the viscous forces  (more information is provided in Appendix ~\ref{app:theory}). The dynamics of the stress-free shape amplitudes $g_{l,m}$'s are assumed to simply rotate with the external flow~\cite{vlahovska2011dynamics,barthes1981time}, dynamically evolving according to
\begin{equation}\label{eq:eom_g}
    \partial_t g_{l,m}= i \omega m g_{l,m} \ .
\end{equation}

The system of equations Eqs.~\eqref{eq:eom}-\eqref{eq:eom_g} is closed by specifying the dynamics of the spatially homogeneous part of the membrane tension $\sigma(t)$. As mentioned, $\Sigma$ acts as a Lagrange multiplier and dynamically evolves to enforce a constant excess area $\Delta$. By enforcing $\partial_t \Delta \!=\! 0$, $\sigma(t)$ is expressed as~\cite{vlahovska2011dynamics}
\begin{equation}\label{eq:st}
    \sigma(t) = -\frac{\mcB}{\sum_{l=2}^\infty \sum_{m=-l}^l\Gamma_2^l a_l f_{l,m} f_{l,m}^*} \left[\sum_{l=2}^\infty \sum_{m=-l}^l a_l \left(\mcB^{-1} \Gamma_1^l f_{l,m}+ \mcC^{-1} \Gamma_3^l f_{l,m}  + S_{l,m}\right)f_{l,m}^*\right] \ .
\end{equation}
This coupled system, Eqs.~\eqref{eq:eom}-\eqref{eq:st}, allows us to explore how the dimensionless numbers $\mcB$, $\mcC$ and $\lambda$, as well as the stress-free and enucleated excess areas, $\Delta_0$ and $\Delta$, influence the system's stability, and eventually its long-time behaviors. While the perturbative approach assumes $\Delta,\Delta_0\!\ll\!1$, we demonstrate the effects below for modest $\Delta$ values --- doing so allows us to emphasize the emerging physics, and to adopt values closer to previous experiments and simulations.

\section{Unperturbed dynamics}\label{se:upd}
Interestingly, the system of Eqs.~\eqref{eq:eom}-\eqref{eq:st} contains a single essential nonlinear contribution --- the coupling of different $f_{l,m}$'s through the membrane tension $\sigma(t)$. Before analyzing the system's dynamics for multiple $l$ values, it is instructive to examine the dynamics in the $l\!=\!2$ subspace only. This reduction, treating the RBC as an ellipsoid, was shown to capture fundamental RBC dynamic regimes such as tumbling, tank-treading, and swinging or vacillating-breathing~\cite{vlahovska2011dynamics}. We demonstrate that the emerging dynamics are determined not only by the membrane elasticity and external flow rate, but also by the reference excess area $\Delta_0$.

\begin{figure}[t]
\includegraphics[width=0.95\textwidth]{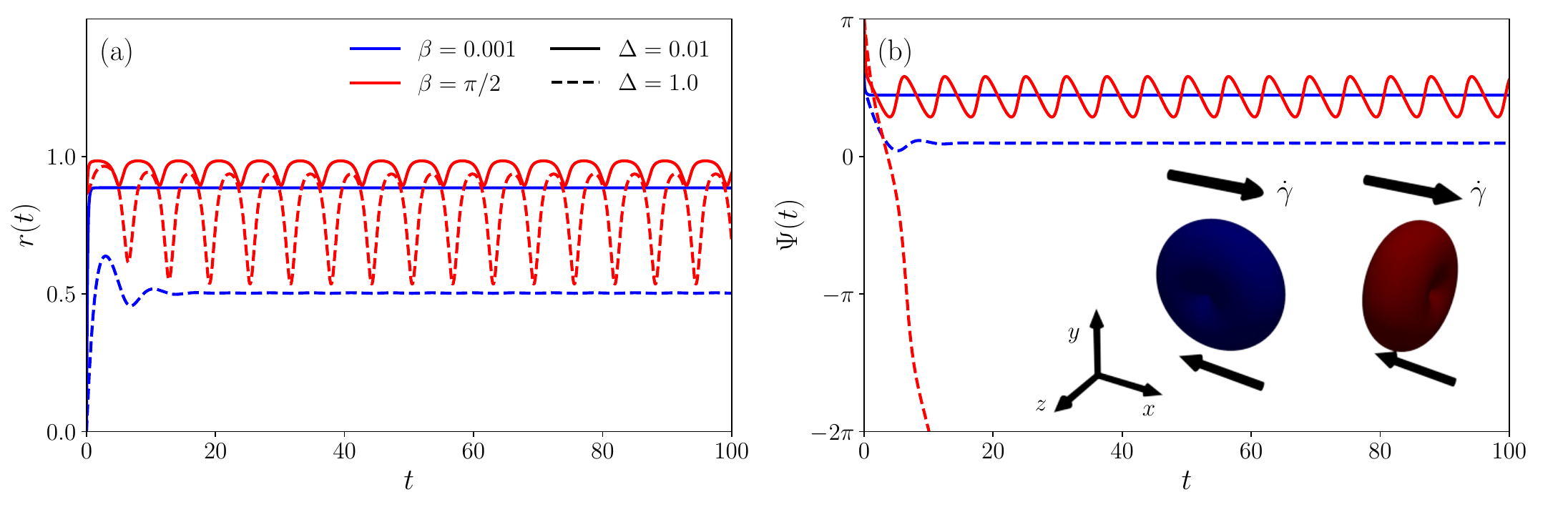}
\caption{Reduced system dynamics with different excess areas. The dynamics of $r$ in (a) and $\Psi$ in (b) for $\mcC\!=\!0.04$, $\lambda\!=\!5$, and $\Delta_0\!=\!0.01$, for $\beta\!=\!0.001$ and $\beta\!=\!\pi/2$, and for different $\Delta$ values $\Delta\!=\!0.01,1$ (corresponding to $S\!\simeq\!2$, and $\Lambda\!\simeq\!0.19, 1.9$ respectively). At long times, the $\beta\!=\!0.001$ dynamics saturate $r$ to a fixed value, such that the RBC is elongated and oriented at a fixed angle $\Psi$ with the external flow. Even though $S$ is not very large, the dynamics converge to the steady states of Eq.~\eqref{eq:ss_red}, as the OP orientation is rotationally symmetric in the shear plane. The $\beta\!=\!\pi/2$ exhibits oscillations --- for $\Delta\!=\!0.01$ the dynamics converge to oscillations around a fixed angle, characteristic of swinging. For $\Delta\!=\!1$ the $r$ oscillations are accompanied by a continuous rotation, visible in the decrease of $\Psi$, which is equivalent to an RBC that deforms and tumbles. Here the solution deviates from the steady-state dynamics of Eq.~\eqref{eq:ss_red}, as $S$ is insufficiently large, and the dynamics are presented over short times. Insets: the OP orientation in blue and the IP orientation in red (here we use $\Delta\!=\!3$ to emphasize the differences).}

\label{fig:fig2}
\end{figure}

We simplify the initial and stress-free structures by assuming they are axially symmetric, concentrating all the excess area in the $(l,m)\!=\!(2,0)$ components of the RBC's stress-free and enucleated structures. To examine the effects of initial orientation, we reorient this initial structure by applying the Wigner D matrix~\cite{rose1995elementary} (a representation of the 3D rotation operator), $D^{(l)}_{m,0}(\beta)$ where $\beta$ is the rotation angle (see Appendix~\ref{app:Ylmrot} for details). That is, we set $f_{2,m}(0)\!=\!-\sqrt{\frac{\Delta}{a_{2}}}D^{(2)}_{m,0}\left(\beta\right)$, and $g_{2,m}(0)\!=\!-\sqrt{\frac{\Delta_0}{a_{2}}}D^{(2)}_{m,0}\left(\beta\right)$. As $D^{(2)}_{2,0}\!=\!\frac{1}{2}\sqrt{\frac{3}{2}}\sin^2\beta$, the resulting initial conditions are symmetric for $f_{2,\pm2}(0)$ and $g_{2,\pm2}(0)$. In what follows, we focus on two extreme cases: $\beta\!=\!0$ corresponding to the RBC's symmetry axis aligned perpendicular to the flow plane, which is an out-of-plane orientation (OP), and $\beta\!=\!\pi/2$ corresponding to an in-plane (IP) symmetry axis, as shown in the inset of Fig.~\ref{fig:fig2}(b). For these two values the $m\!=\!\pm1$ components vanish.

We analytically reduce the equations of motion by parameterizing $f_{2,\pm2}$, $f_{2,0}$, and $g_{2,\pm2}$, $g_{2,0}$ as~\cite{ misbah2006vacillating,vlahovska2011dynamics}
\begin{subequations} \label{eq:subs}
    \begin{align}
f_{2,\pm2}(t)\!=\!\sqrt{\frac{\Delta}{2 a_2}} r(t) e^{\mp 2 i \psi(t)} & \quad \ ,\quad  f_{2,0}(t)\!=\!-\sqrt{\frac{\Delta}{a_{2}}\left[1-r(t)^{2}\right]} \ , \\
   g_{2,\pm2}(t)\!=\!\sqrt{\frac{\Delta_0}{2 a_2}} r_0(t) e^{\mp 2 i \phi(t)} &\quad  \ , \quad g_{2,0}(t)\!=\!-\sqrt{\frac{\Delta_0}{a_{2}}\left[1-r^{2}_0(t)\right]} \ ,
    \end{align}
\end{subequations}
where the variables $r(t)$ and $r_0(t)$ represent, respectively, the deformation amplitudes in the shear plane of the current and stress-free shapes, while the angles $\psi(t)$ and $\phi(t)$ represent the respective in-plane orientations.
The $f_{2,0}$ and $g_{2,0}$ components are obtained by the conservation of the excess areas $\Delta$ and $\Delta_0$, respectively. Finally, we chose $f_{2,0}$ and $g_{2,0}$ with a negative sign consistent with the emerging oblate shape of RBCs.


Substituting these parameterizations into the membrane tension $\sigma(t)$ of Eq.~\eqref{eq:st} yields the reduced membrane tension $\sigma_R(t)$ as
\begin{equation}\label{eq:st_red}
\sigma_R(t)=-6+\frac{\sqrt{30\pi}\mathcal{B}}{3\sqrt{\Delta}}r\sin\Psi+\frac{2\mathcal{B}}{3\mathcal{C}}\sqrt{\frac{\Delta_0}{\Delta}}\left[r_{0}r\cos(\Phi-\Psi)+\sqrt{\left(r_{0}^{2}-1\right)\left(r^{2}-1\right)}-\sqrt{\frac{\Delta}{\Delta_0}}\right] \ ,
\end{equation}
where we defined $\Psi\!\equiv\!2\psi$ and $\Phi\!\equiv\!2\phi$, and suppressed the explicit time dependence from $r$, $r_0$, $\Psi$ and $\Phi$ for readability.
Inserting the parameterizations Eq.~\eqref{eq:subs} together with Eq.~\eqref{eq:st_red} into the equations of motion Eqs.~\eqref{eq:eom}-\eqref{eq:eom_g} yields reduced dynamical equations for $r$, $\Psi$ and $\Phi$ as~\cite{vlahovska2011dynamics}
\begin{subequations}\label{eq:red_eqs}
    \begin{align}
        r' =& \Lambda^{-1}\left(1-r^{2}\right)\sin\Psi+\left(\Lambda S\right)^{-1}\left[r_0\left(1-r^{2}\right)\cos(\Phi-\Psi)-r\sqrt{\left(1-r_0^{2}\right)\left(1-r^{2}\right)}\right] \ , \label{eq:r_eq_unprt}\\
        \Psi' =& -1+\Lambda^{-1}\frac{\cos\Psi}{r}+\left(\Lambda S\right)^{-1}\frac{r_{0}}{r}\sin(\Phi-\Psi) \ , \label{eq:psi_eq_unprt} \\
        \Phi' =& -1 \ ,
    \end{align}
\end{subequations}
where $\bullet'\equiv \tfrac{d\bullet}{dt}$ is the time-derivative operator. The dynamics are governed by two composite dimensionless parameters 
\begin{equation} \label{eq:Lambda_S}
\Lambda\!=\!\frac{\sqrt{\Delta } (23 \lambda +32)}{8 \sqrt{30 \pi }} \! \qquad \ , \qquad S\!=\!\mcC \sqrt{\frac{15 \pi }{2 \Delta_0}} \ ,
\end{equation}
and the initial value $r_0$. Clearly, this reduced system depends on the combined effects of the viscosity ratio $\lambda$, the capillary number $\mcC$, and the excess areas $\Delta$ and $\Delta_0$. The equations above are fully determined once $\Lambda$, $S$ and $r_0$ are specified together with the initial condition characterized by $\beta$. Interestingly, while the bending number $\mcB$ appears in the reduced membrane tension Eq.~\eqref{eq:st_red}, it is absent from the equations of motion of the reduced system, Eq.~\eqref{eq:red_eqs} --- this is due to the fact that $\sigma_R$ is determined only using the $l\!=\!2$ components. Previous analysis~\cite{vlahovska2011dynamics} with $\Delta\!=\!\Delta_0$ demonstrated Eq.~\eqref{eq:st_red} can exhibit various dynamical behaviors, e.g., fluid membrane and capsule dynamics.  As recent experimental evidence suggests a shape instability at  intermediate and large shear rates, we are interested in the solutions for large $\mcC$, and consequently large $S$ values. In the limit $S\!\gg\!1$, the leading order steady-state solutions are obtained as~\cite{vlahovska2011dynamics}
\begin{subequations} \label{eq:ss_red}
    \begin{alignat}{3}
r_{ss}^{-} & = 1 \quad , \qquad & \Psi_{ss}^{-} & =\arccos\Lambda \quad , & \quad \text{for }\Lambda<1 , \label{eq:ss_red_TT} \\
r_{ss}^{+} & = \Lambda^{-1} \quad , \qquad & \Psi_{ss}^{+} & =0 \quad , & \quad \text{for }\Lambda>1  , \label{eq:ss_red_VB}
\end{alignat}
\end{subequations}
where we used the notation $\bullet^-_{ss}$ for the $\Lambda\!<\!1$ steady-state solutions, and $\bullet^+_{ss}$ for the $\Lambda\!>\!1$ ones.

Unlike in previous analysis~\cite{vlahovska2011dynamics}, $S$ defined in Eq.~\eqref{eq:Lambda_S} incorporates the reference shape excess area $\Delta_0$, while $\Lambda$ depends on $\Delta$. $S$ then becomes the effective elastic constant, quantifying the ability of the shear flow to overcome the energy barrier associated with the tank-treading of the heterogeneous membrane~\cite{dupire2015simple, mendez2018plane}. This modified definition of $S$ implies that a dynamical transition can occur either by increasing the shear rate through $\mcC$, or by decreasing $\Delta_0$. In the latter case, the reference shape gets close to a sphere, reducing the heterogeneity of the reference shape, which facilitates its circulation and favors tank-treading,
consistent with previous studies~\cite{dupire2015simple, mendez2018plane, peng2014erythrocyte, abkarian2007swinging}. Clearly, such a dynamical transition is unavailable for vesicles, as their fluidized membrane does not hold memory of a previous state. In our framework, this corresponds to either $\Delta_0\!\rightarrow\!0$ or $\mcC\!\rightarrow\!\infty$ (see Appendix~\ref{app:vesicle}).

We briefly examine the dynamics of Eq.~\eqref{eq:red_eqs} as a function of the initial orientations and the excess areas $\Delta$ and $\Delta_0$. The initial orientation is controlled by the angle $\beta$ and results in different values of the initial deformation amplitude $r(0)$. We assume that the initial orientation of the current shape and the stress-free shape are the same. Ideally, we would initialize an OP RBC (with $\beta\!=\!0$), and then apply rotations with $D^{(2)}_{\pm2,0}\left(\beta\right)$ to both $f_{2,m}$'s and $g_{2,m}$'s to explore the role of initial orientation. Practically, setting the stress-free and initial structures in an OP orientation ($\beta\!=\!0$) results in $r\!=\!0$, which causes a singularity of Eq.~\eqref{eq:red_eqs}. To avoid this singularity, we impose a small initial rotation $\beta\!\simeq\!0$ (rendering the $f_{2,\pm1}$ components negligible). To obtain the IP case we apply $D^{(2)}_{\pm2,0}\left(\pi/2\right)$ to $f_{2,m}$'s and $g_{2,m}$'s (yielding $f_{2,\pm2}\!=\!-\frac{1}{2}\sqrt{\frac{3 \Delta}{2 a_2}}$, $f_{2,0}\!=\!\frac{1}{2}\sqrt{\frac{\Delta}{a_2}}$ and similar values for $g_{2,\pm2}$ and $g_{2,0}$ upon replacing $\Delta$ with $\Delta_0$, giving $r(0)\!=\!r_0(0)\!=\sqrt{3}/2$, and $\Psi(0)\!=\!\Phi(0)\!=\!\pi$). 

We analyze the dynamics for different excess areas $\Delta$ in Fig.~\ref{fig:fig2}, fixing $\lambda\!=\!5$, a typical value for physiological conditions~\cite{wells1969red,lanotte2016red}. We probed different $\Delta$ values to demonstrate the effect of internal stresses, using $\Delta\!=\!\Delta_0\!=\!0.01$ as a reference, non-stressed case. For both the OP $\beta\!\simeq\!0$ and IP $\beta\!=\!\pi/2$ orientations, increasing $\Delta$ lowers the $r$ value at long times, indicating a less elongated particle in the shear plane. In the OP case of $\beta\!\simeq\!0$, the emerging RBC dynamics correspond to RBC alignment along the vorticity axis with a fixed deformation and orientation. This motion is actually accompanied by the tank treading of the membrane. This tank-treading motion differs from the classical case~\cite{fischer1978red} associated with an IP orientation. In the present case, the motion can be interpreted as a deformed form of rolling~\cite{dupire2012full}: the RBC’s short axis is aligned with the vorticity axis, the membrane tank-treads around the cell’s short axis, while shear stresses deform the cell within the shear plane~\cite{yao2001low} and tend to align it with the flow direction ($\Psi$ decreases), as interpreted from the blue curves in Fig.~\ref{fig:fig2}(a)-(b). In the IP case of $\beta\!=\!\pi/2$, the heterogeneity of the reference shape makes the particle shape and angle oscillate over time~\cite{abkarian2007swinging}. Increasing $\Delta$ promotes a transition from that swinging motion to tumbling, visible in Fig.~\ref{fig:fig2}(b). As changing $\Delta$ amounts to changing $\Lambda$, similar transitions have been observed for $\Delta\!=\!\Delta_0$ upon increasing $\lambda$~\cite{keller1982motion,vlahovska2011dynamics, kaoui2009vesicles}.

The enucleation process allows the RBCs to lock stresses within their membranes, manifested by an excess area $\Delta$ different from their reference value $\Delta_0$. The effective elasticity constant $S$ is then renormalized, taking into account the reference excess area $\Delta_0$ instead of $\Delta$. 
Additional examples of the dynamics available for the system described above are available in Appendix~\ref{app:vesicle}. While the reduced system successfully captures various RBC dynamics, it is inherently limited to the ellipsoidal modes, and hence cannot provide insight into the possible reshaping of RBCs. This motivates us to include and analyze the dynamics in the presence of higher-order spherical harmonics. In what follows we demonstrate that enucleation, and the mismatch $\Delta\neq\Delta_0$, not only modify the unperturbed dynamics, but also allow for shape instabilities to emerge.

\section{Perturbations and instabilities}\label{se:inst}

Understanding the reduced system's dynamics, we now return to the full perturbative formalism of Eqs.~\eqref{eq:eom}-\eqref{eq:st}, including contributions from higher-order modes, $l\!>\!2$. 
In this section, we start by deriving the stability criterion to clarify the role of the different parameters of the problem. We then use the reduced dynamics to obtain quantitative instability predictions, which we then verify numerically by integrating Eqs.~\eqref{eq:eom}-\eqref{eq:st}. 

According to Eqs.~\eqref{eq:eom} and~\eqref{eq:eom_sup}, an instability is expected once the real part of the growth rate becomes positive $\Re\left[\Upsilon_l\right]\!>\!0$ for any mode $l\!>\!2$. Using the expression for $\Upsilon_l$ in Eq.~\eqref{eq:eom_sup_ups}, we can recast this set of instability criteria per $l$-mode, as a simplified condition on the instantaneous membrane tension
\begin{equation} \label{eq:inst}
    \Pi^2 + \sigma(t) \Pi + 4 \alpha < 0 \ ,
\end{equation}
where we defined $\Pi\!\equiv\!l(l+1)$ for convenience. The above simplified criterion also introduces the dimensionless ratio between the in-plane shear elasticity to the bending stiffness, $\alpha\!\equiv\!\mcB/\mcC\!=\!\mu R^2 / \kappa$, reminiscent of the F\"oppl–von K\'arm\'an number~\cite{fedosov2014multiscale,AudolyPomeau2010}, but for composite membranes. 
An instability occurs once the bending and elastic responses are insufficient to counter the forces induced by the external flow and internal stresses. A similar instability criterion was obtained for vesicles under external drive~\cite{turitsyn2008wrinkling}.

When $\sigma(t)$ is sufficiently negative, the system is unstable. We treat $l$ as a continuous variable, resulting in unstable, excited $l$-modes existing between $l_{\pm}\!=\!\frac{1}{2} \left(-1+\sqrt{1-2 \sigma\pm 2 \sqrt{\sigma ^2-16 \alpha }}\right)$, with a maximum at $l_c\!=\!\frac{1}{2}\left(-1+\sqrt{1-2\sigma}\right)$. At the critical point, we set the inequality of Eq.~\eqref{eq:inst} to equality, allowing us to isolate $\sigma(t)\!=\!-\frac{4\alpha+\Pi^2}{\Pi}$. The maximal $l$ value then yields $l_*\!=\!\frac{1}{2}\left(-1+\sqrt{1+8\sqrt{\alpha}}\right)$, and the corresponding $\sigma_*\!\equiv\!-4\sqrt{\alpha}$. Importantly, as physically only integer $l$ values are allowed, even if the instability criterion is met, we need to ensure it also holds for a subset of integer values within $l_- \le l \le l_+$.

\begin{figure}[t]
\includegraphics[width=0.95\textwidth]{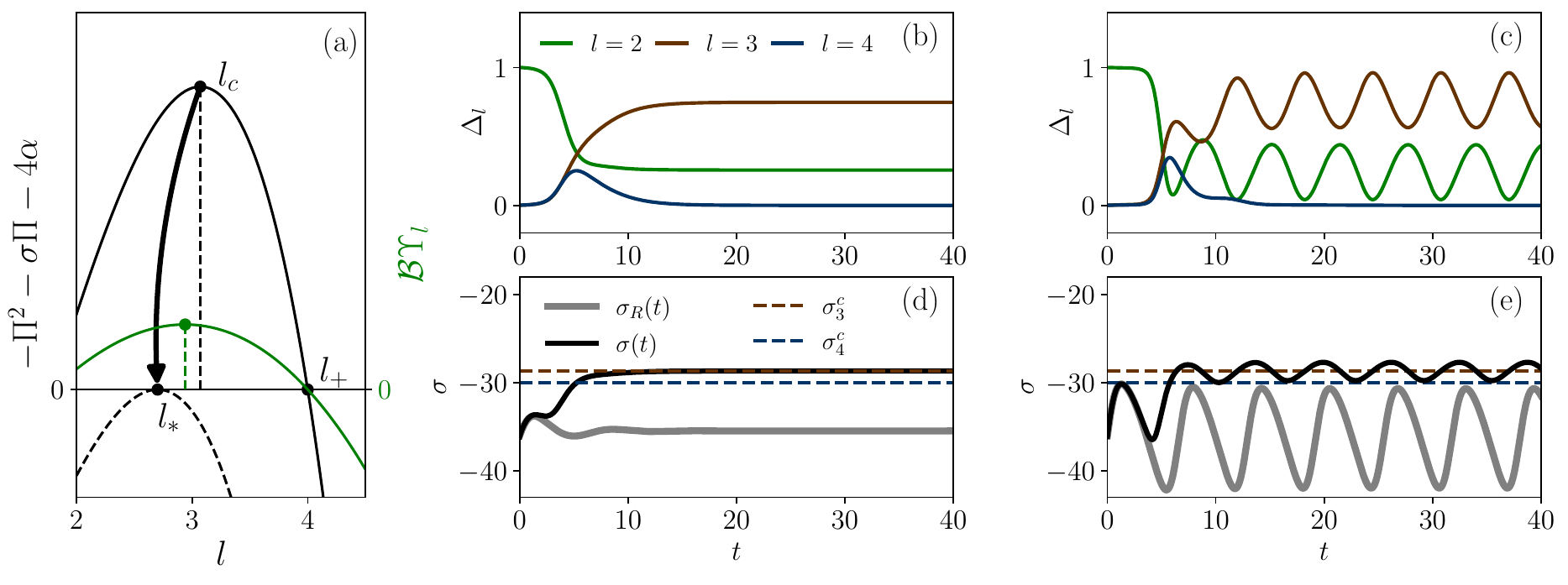}
\caption{Linear instability criteria and dynamics. (a) The simplified criterion $-\Pi^2 - \sigma(t) \Pi - 4 \alpha \! >\! 0$ [note the negative sign compared to Eq.~\eqref{eq:inst}], evaluated for $\alpha\!=\!25$, at the critical stress $\sigma(t)\!=\sigma_*\!\equiv\!-4\sqrt{\alpha}$ (denoted by a dashed line), and for $\sigma(t)\!<\!\sigma_*$ (denoted by a solid black line). Positive values correspond to unstable modes, while negative values are stable modes. Black points represent $l_*$ and $l_c$ (connected by a black arrow), as well as the root $l_+$. We also plot in green the corresponding curve for $\mcB \Upsilon_l$ for $\sigma(t)\!<\!\sigma_*$, as a function of $l$ (using the same $\alpha$ and $\sigma$ values, but on a different vertical scale), to demonstrate that while the roots are identical to the simplified expression, $l_c$ can differ from the fastest-growing mode (vertical black and green dashed lines serve as a guide to the eye). The bold curved arrow indicates that initially-unstable dynamics cause $\sigma$ to increase, and the stability curve dynamically evolves from the solid, unstable curve towards the marginally-stable, dashed curve. (b)-(c) $\Delta_l$ dynamics for $\mcC\!=\!0.04$, $\mcB\!=\!2$, $\Delta\!=\!1$, $\Delta_0\!=\!0.01$, $\lambda\!=\!5$ for $\beta\!=\!0.001$ in (b) and $\beta\!=\!\pi/2$ in (c). In both cases the initial ellipsoid transitions the excess area from $\Delta_2$ to $\Delta_3$ and $\Delta_4$, until $l\!=\!4$ is stabilized, and $\Delta_4$ starts decreasing. The emerging dynamics are reminiscent to those of the reduced system --- smooth trajectories for $\beta\!\simeq\!0$ and oscillatory ones for $\beta\!=\!\pi/2$ (see Fig.~\ref{fig:fig2}). (d)-(e) The temporal dynamics of the membrane tension $\sigma(t)$ in black, the reduced system's membrane tension $\sigma_R(t)$ in thick gray, and $\sigma_l^c$ evaluated for $l\!=\!3,\!4$. For the OP case in (d), $\sigma(t)$ initially follows $\sigma_R(t)$ but is below $\sigma_3^c$ and $\sigma_4^c$, which causes both to increase and $\Delta_2$ to decrease. Eventually $\sigma\!>\!\sigma_4^c$, causing $\Delta_4$ to decrease. In (e) the period-averaged $\sigma(t)$ follows similar dynamics.}
\label{fig:fig3}
\end{figure}

The theoretical criterion allows us to predict whether a certain parameter combination is stable or not, and also which modes are expected to grow the most. An instability is initiated when $\sigma\!<\!\sigma_*$, corresponding to the solid curve in Fig.~\ref{fig:fig3}(a). The growth of the excited modes causes an increase in $\sigma$ towards $\sigma_*$ that stabilizes the unstable modes. This increase in $\sigma$ pushes the stability curve, from the unstable curve [corresponding to the solid line in Fig.~\ref{fig:fig3}(a)], towards the marginally-stable curve [corresponding to the dashed line in Fig.~\ref{fig:fig3}(a)], as is shown schematically by the bold arrow in Fig.~\ref{fig:fig3}(a). As the dynamics evolve, some of the modes are stabilized, and eventually the modes that are unstable for the longest times correspond to $l_*$. Finally, as the instability criterion differs from the growth rates $\Upsilon_l$ by an $l$-dependent factor, $l_c$ is not necessarily the fastest-growing mode. We schematically visualize $l_c$, $l_*$, $l_{+}$, the instability criterion $\sigma(t)\!=\!\sigma_*$, and the rates $\mcB \Upsilon_l$ in Fig.~\ref{fig:fig3}(a), showing explicitly a possible deviations of $l_c$ from the instantaneous fastest growing mode.

To induce a shape change, the excess area needs to be redistributed between $l\!=\!2$ and other $l$'s. We demand that the instability occurs for $l\!\geq\!3$, which implies that $l_+\!>\!3$.  We simplify this demand by assuming that the unstable range in $l$ is approximately symmetric, which allows us to check whether $l_*\!\ge\!2.5$ instead. As this condition is time independent and depends solely on $\alpha$, it implies $\alpha\!\ge\!\left(35/8\right)^2$. 

Predicting the instability onset criteria analytically is challenging, because the equations of motion are nonlinearly coupled through $\sigma(t)$. To advance, we utilize our assumption that the initial RBC shape is an ellipsoid, so the RBC's excess area is initially concentrated in $l\!=\!2$ (except from potential perturbations), allowing us to approximate $\sigma(t)\!\simeq\!\sigma_R(t)$ of Eq.~\eqref{eq:st_red}. As long as the dynamics are stable, the only significant shape amplitudes remain within the $l\!=\!2$ subspace, and the system evolves according to the reduced description of Eq.~\eqref{eq:red_eqs}. For an instability to occur $\sigma(t)\!\simeq\!\sigma_R(t)\!<\!\sigma_*$. In general, this non-trivial evolution links $\mcB$, $\mcC$, $\Delta_0$, $\Delta$ and $\lambda$, as well as the initial conditions. To gain some intuition for this criterion, we can estimate $\sigma(t)$ by evaluating $\sigma_R(t)$ at convenient states --- the initial conditions or the long-time behaviors for $\Lambda\!<\!1$ and $\Lambda\!>\!1$ of Eq.~\eqref{eq:ss_red}. 

We first use the OP orientation and set $r,r_0\!\rightarrow\!0$ in Eq.~\eqref{eq:st_red}, which render $\Psi$ and $\Phi$ irrelevant. The initial value of $\sigma_R(t)$ is approximated as $\sigma_R^{\text{init}}\!\simeq\! -6+\frac{2}{3}\alpha\left(\sqrt{\Delta_0 / \Delta}-1\right)$, and demanding that this evaluates to $\sigma_*$ yields a critical ratio $\alpha_*^{\text{init}}\!=\!
9/\left[1-\left(\Delta_0 / \Delta\right)^{1/4}\right]^2$. An instability should satisfy both $l_*\!\ge\!2.5$ [corresponding to $\alpha\!\ge\!(35/8)^2$], and $\alpha\!>\!\alpha_*^{\text{init}}$. With $\Delta_0\!=\!0.01$ and $\Delta\!=\!1$ the two values are similar, and we predict an instability once both of these criteria are met.

In case the underlying reduced dynamics evolves quickly towards their long-time dynamics, we approximate $\sigma_R(t)$ in a similar fashion, using the long-time solution for the reduced system of Eq.~\eqref{eq:red_eqs}. Using the steady-state approximate solutions Eq.~\eqref{eq:ss_red} in $\sigma_R(t)$, we equate the time-averaged membrane tension $\langle \sigma_R^{\bullet} \rangle$ to $\sigma_*$ (as detailed in Appendix~\ref{app:reducedsteadystates}). For the $\Lambda\!<\!1$ and $\Lambda\!>\!1$ scenarios in the limit of weak elasticity $S\!\gg\!1$, we obtain the critical $\mcC$ values as
\begin{equation}\label{eq:c_crit}
    \mcC^{-}_* = \frac{2 }{3\left(\sqrt{6}+\sqrt{\mcB A^{-}}\right)^2} \mcB\ , \qquad
    \mcC^{+}_* = \frac{1}{9} \left(1-\sqrt{A^{+}}\right)^2\mcB\ ,
\end{equation}
where we defined the auxiliary $A^{-}\!\equiv\!\frac{1}{3}\sqrt{\frac{30 \pi (1-\Lambda^2)}{\Delta}}$, and $A^{+}\!\equiv\!\sqrt{\frac{\Delta_0}{\Delta}\frac{1-\Lambda^2}{\Lambda}\left(r_0^2-1\right)}$ respectively. While the $\Lambda\!>\!1$ solution depends trivially on $\mcB$ (i.e., the instability criterion could be cast in terms of $\alpha$), the $\Lambda\!<\!1$ case results in a non-trivial critical condition. 

To test these predictions, we numerically integrate Eqs.~\eqref{eq:eom}-\eqref{eq:st} for $l\!\in\!\left[2,10\right]$, and $m\!\in\!\left[-l,l\right]$. We initialize the $f_{l,m}$'s and $g_{l,m}$'s by rotating a fixed rest structure, as described previously (see also Appendix~\ref{app:Ylmrot}). For convenience, we first define the excess area stored in the $l$-th subspace as $\Delta_l\!\equiv\!a_l \sum_{m=-l}^l f_{l,m}f_{l,m}^*$. We choose to start with $f_{2,m}$'s and $g_{2,m}$'s as used in Sec.~\ref{se:upd}, that is we incorporate $\Delta_2\!=\!\Delta$ and $\Delta_0$ in the axial direction $f_{2,0}$ and $g_{2,0}$ respectively. To test the stability of higher modes, we introduce small initial perturbations of amplitude $\delta$ to $f_{l,0}$ with $l\!>\!2$, such that $\Delta_l\!=\!\delta$ for $l\!\in\!\left[3,10\right]$ (here we chose $\delta\!=\!10^{-3}$), while we keep the stress-free shape purely ellipsoidal, setting $g_{l,m}\!=\!0$ for all $m$ and $l\!>\!2$. By considering such axial perturbations, we assume that the enucleation process preserves the planar symmetry of the RBC but breaks the top-bottom reflection symmetry. To rotate this shape, we apply $D^{(l)}_{m,0}\left(\beta\right)$ to $f_{l,0}$ and $g_{l,0}$ (e.g., set $\beta\!=\!10^{-3}$ for the OP case, or $\beta\!=\!\frac{\pi}{2}$ for the IP case). We then integrate Eqs.~\eqref{eq:eom}-\eqref{eq:st} for $t\!\in\!\left[0,T\right]$ (we take $T\!=\!500$ in what follows).

We show two examples of the dynamics obtained for $\Delta\!>\!\Delta_0$ in Fig.~\ref{fig:fig3}(b)-(e), with similar parameter values as used in Fig.~\ref{fig:fig2}. Panels (b)-(c) show the dynamics of $\Delta_l$ for $l\!=\!2\!-\!4$, for the OP and IP cases respectively. In both cases, $\Delta_2$ decreases while $\Delta_3$ and $\Delta_4$ increase initially. Eventually, $\Delta_4$ decreases while $\Delta_3$ continues to increase until the system reaches a steady-state or oscillates around such a state. The redistribution of $\Delta$ implies the RBC changes its shape. Panels (d)-(e) show the corresponding membrane tension $\sigma(t)$ obtained from the full dynamics in black, as well as the reduced membrane tension $\sigma_R(t)$ [obtained from the reduced system, Eq.~\eqref{eq:st_red}] in gray. On top of the $\sigma(t)$ dynamics, we also plot in dashed lines the instability criterion $\sigma_l^c$ evaluated for different $l$ values. The initial dynamics of $\sigma(t)$ closely follows $\sigma_R(t)$ while an instability develops, causing $\sigma(t)$ to increase. As $\sigma(t)\!<\!\sigma_3^c,\sigma_4^c$ initially, both $\Delta_3$ and $\Delta_4$ grow. This causes $\sigma(t)$ to grow until $\sigma_4^c\!<\sigma(t)$, causing $\Delta_4$ to decrease while $\Delta_3$ keeps increasing ($\Delta_2$ decreases to conserve the total excess area). Eventually $\sigma(t)\!=\!\sigma_3^c$, stabilizing all growth in the system. While the initiation of the instability happens for both OP and IP orientations, the emerging dynamics are rather different; this results from the fact that the instability criterion could be cast in terms of $\Delta_l$, which is agnostic to the RBC orientation (see Appendix~\ref{app:theory}). 

\begin{figure}[t]
\includegraphics[width=0.97\textwidth]{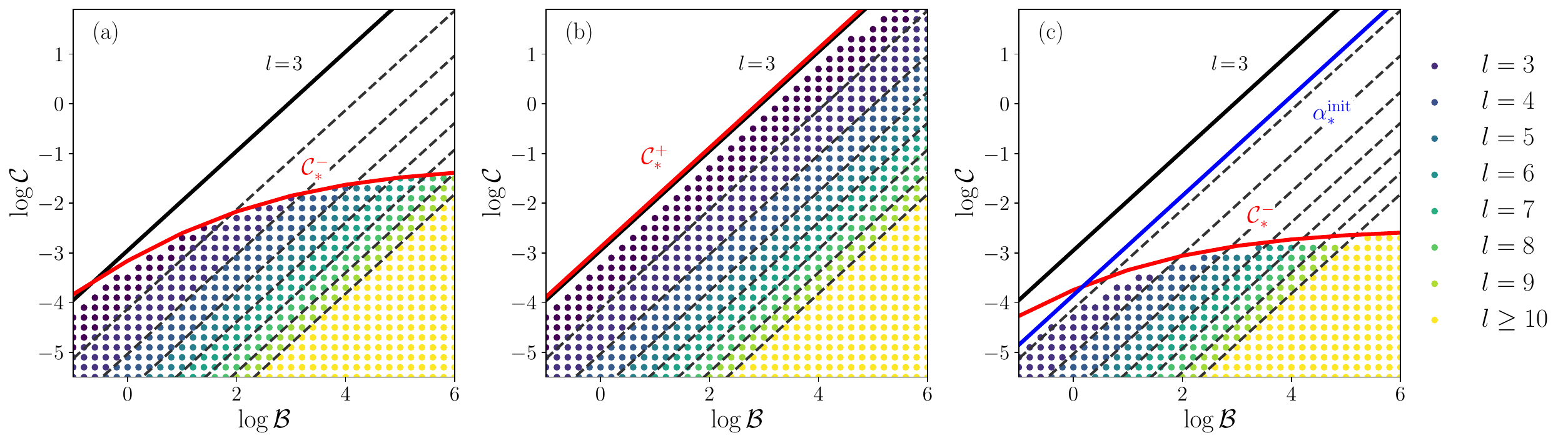}
\caption{Out of plane (OP) stability phase diagrams for different $\lambda$ values [panels (a)-(b)], and different $\Delta$ values [panels (b)-(c)]. $\log \mcB-\log\mcC$ phase diagrams for (a) $\lambda\!=\!1$, $\Delta\!=\!1$ ($\Lambda\!\simeq\!0.7$), (b) $\lambda\!=\!5$, $\Delta\!=\!1$ ($\Lambda\!\simeq\!1.9$), and (c) $\lambda\!=\!5$, $\Delta\!=\!0.1$ ($\Lambda\!\simeq\!0.6$) in the OP orientation ($\Delta_0\!=\!0.01$ for all cases). Each phase diagram was obtained by sampling $\left(\mcB,\mcC\right)$ pairs spanning $\mcB\!\in\!\left[\exp(-1),\exp(6)\right]$, and $\mcC\!\in\!\left[\exp(-5.5), \exp(2)\right]$, and simulating each perturbed system for a duration of $T\!=\!500$. Colors indicate the value of $l_\Delta$, signifying the $l$ value for which the excess area increased the most (and  empty regions correspond to stable dynamics with $l_{\Delta}\!=\!2$), as obtained from the numerical simulations (see legend). Solid diagonal black lines show the theoretical predictions for $l_*\!=\!2.5$ (where we expect $l\!=\!3$ to be excited), and the additional diagonal dashed lines show the theoretically predicted transition from excited $l$ value to the next one. Red curves show the value of $\mcC_*$ obtained from the steady-state calculation [Eq.~\eqref{eq:c_crit}, see also Appendix~\ref{app:reducedsteadystates}]. Finally, the blue diagonal line shows $\alpha_*^{\text{init}}$ effecting the instability boundary at low $\mcB$ and $\mcC$ values [$\alpha_c^{\text{init}}$ is not shown in panels (a) and (b) as it roughly overlaps with the $l\!=\!3$ critical line, as mentioned above]. Increasing the viscosity ratio $\lambda$, and the excess area $\Delta$ [panel (b) contrasted with (a) and (c)]  destabilizes RBCs of high $\mcC$ (weak elasticity).}
\label{fig:fig4}
\end{figure}

\begin{figure}[h]
\includegraphics[width=0.75\textwidth]{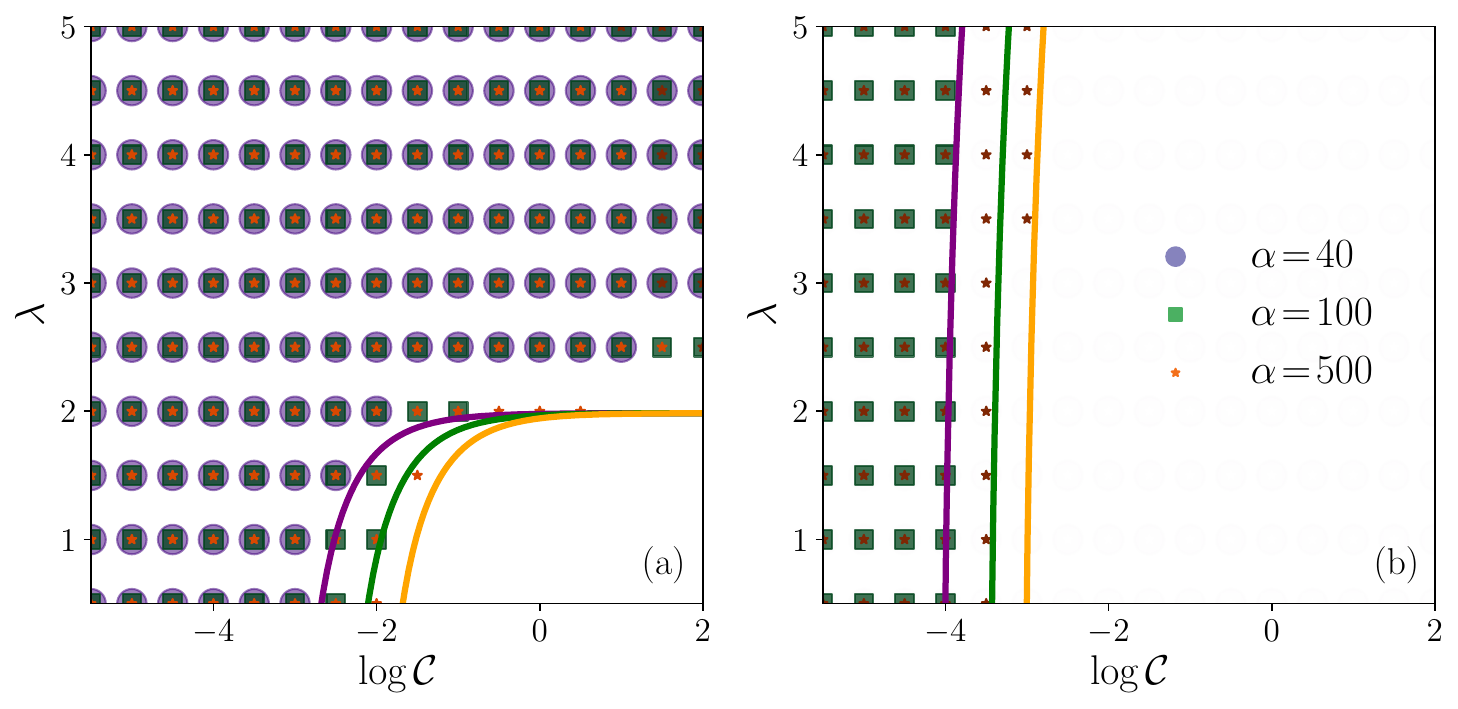}
\caption{Out of plane (OP) stability diagrams in $\log \mcC-\lambda$ space, obtained for $\alpha\!=\!40,100$ and $500$ for $\Delta\!=\!1$ in (a) and for $\Delta\!=\!0.1$ in (b). Markers represent unstable dynamics for the associated $\alpha$ value (and  empty regions correspond to stable dynamics). We plot the predictions for $\lambda_*^{-}(\mcC)$ of Eq.~\eqref{eq:ss_lambda_red} in both panels for the three $\alpha$ values (see Appendix~\ref{app:reducedsteadystates}). Increasing $\alpha$ pushes the boundary to higher $\mcC$ values; a similar effect happens for increased $\Delta$ values. Note that in panel (b), $\alpha\!=\!40$ is actually below the $\alpha_*^{\text{init}}$ criterion, rendering the entire $\log \mcC-\lambda$ space stable (we still plot the theoretical prediction $\lambda_*^{-}$). The $\log \mcC-\lambda$ phase diagrams are consistent with the observations from Fig.~\ref{fig:fig4} --- increasing $\lambda$ destabilizes RBCs with large $\mcC$.}
\label{fig:fig5}
\end{figure}

To validate the predictions of the stability boundary, we focus on the OP configuration, and systematically sample the $\mcB-\mcC$ space. 
For each case, we numerically find which $l$ mode corresponds to the largest excess area increase, by comparing the initial $\Delta_l(0)$ distribution to the final state $\Delta_l(T)$'s and finding the $l$ value corresponding to the largest $\Delta_l$ growth, $l_\Delta\!=\!\text{argmax}_l \left[\Delta_l(T) - \Delta_l(0)\right]$. We then use $l_\Delta$ values to color the emerging $\left(\mcB,\mcC\right)$ phase space. We sample such phase spaces for $\lambda\!=\!1$ and $\lambda\!=\!5$ for different $\Delta$ values, and plot the results for the OP cases in Fig.~\ref{fig:fig4}. On top, we overlay the theoretically-predicted $\alpha$ values that correspond to different $l_*$ values, resulting in the dashed diagonal lines in Fig.~\ref{fig:fig4}. We also plot the approximate $\mcC_*$ relevant for the specific cases [Eq.~\eqref{eq:c_crit}], and $\alpha_*^{\text{init}}$ when relevant.

The theoretical predictions are verified numerically. Figure ~\ref{fig:fig4}(a) shows the $\lambda\!=\!1$, $\Delta\!=\!1$ case, corresponding to $\Lambda\!<\!1$. Upon introducing the different excess areas we observe an instability in a large region of the $\left(\mcB,\mcC\right)$ phase diagram. The critical $\mcC_*$ value plays a dominant role in stabilizing high $\mcC$ cases. This stabilizing effect is diminished as we increase $\lambda$ to $5$ in panel (b), for which $\Lambda\!>\!1$. Finally, to examine the stabilizing effect of $\alpha_*^{\text{init}}$, we reduce $\Delta$ to $0.1$ (so that $\alpha_*^{\text{init}}$ deviates from the $l_*\!\ge\!2.5$ criterion) in panel (c) --- the stabilizing effect of $\alpha_*^{\text{init}}$ is observed for low $\mcB$-$\mcC$ values. 

In general, the value of $\alpha$ corresponds well to the emerging maximally-unstable observed $l$, implying that the mechanical ratio $\alpha$ dictates the emerging excited harmonics. Additionally, comparing the different panels demonstrates clearly that lower viscosity ratios $\lambda$ and lower excess areas $\Delta$ both stabilize the dynamics for large $\mcC$ values. The stabilizing effects of low $\lambda$ values, or lower $\Delta$ values, are in fact expected, as the underlying dynamics is governed by the reduced equations, Eq.~\eqref{eq:red_eqs}, which depend on the composite parameter $\Lambda$. Similar effects occur for the IP case --- we discuss and compare those to the OP cases in Appendix~\ref{app:in_and_out}.

We also probed stability as a function of $\mcC$ and $\lambda$ (as probed in~\cite{mauer2018flow}). We probed the $\mcC\!-\!\lambda$ phase space for  fixed $\alpha$ values, $\alpha\!=\!40,100$ and $\alpha\!=\!500$ for the OP case. As a fixed $\alpha$ value always induces an instability for the same $l$ value, we only probed whether the dynamics are stable or not. We also derive an analogue instability criteria $\lambda_*(\mcC)$ for fixed $\alpha$ for the $\Lambda\!<\!1$ case, as
\begin{equation} \label{eq:ss_lambda_red}
     \lambda_*^{-} = \frac{8}{23}\left(\frac{\sqrt{\left(30 \pi \mcC^2 \alpha^2 / \Delta\right)-4\left(\sqrt{\alpha}-3\right)^4}}{\mcC \alpha} - 4\right) \ . 
\end{equation}
In the $\Lambda\!>\!1$ case, $\lambda_*$ is independent of $\mcC$ (see Appendix~\ref{app:reducedsteadystates}). We plot the analytical predictions on top of the numerical results for $\Delta\!=\!1$ in Fig.~\ref{fig:fig5}(a) and $\Delta\!=\!0.1$ in Fig.~\ref{fig:fig5}(b). As expected, regions of large $\mcC$ and low $\lambda$ values are stable (see Fig.~\ref{fig:fig4}). This stabilization at high $\mcC$ and low $\lambda$ is reminiscent of the numerically-obtained phase space reported in~\cite{mauer2018flow}. However, the current theoretical framework induces instabilities for fixed $l$ values, which differs from the numerical results of~\cite{mauer2018flow}, hinting that a different instability mechanism may be relevant for the full numerical simulations. 

The observed diagrams imply that RBC become unstable for large $\alpha$ values, which is an intrinsic structural property. If $\Lambda$ is sufficiently low (either due to a lower viscosity ratio $\lambda$ or due to lower excess area $\Delta$), the structural criterion is further stabilized [as seen in Fig.~\ref{fig:fig4}(a) and (c)]. Irrespective of the value of $\Lambda$, the instabilities persist for weak flows associated with low $\mcB$ and $\mcC$ values. This implies that even in the presence of weak flows, RBCs can become unstable. This behavior is somewhat counter-intuitive, as in the absence of flow the RBC should be stable (this is an underlying assumption about the initial RBC state). However previous studies of depression of spherical capsules and vesicles have also reported reminiscent phenomena~\cite{quilliet2006depressions,quemeneur2012gel}. The derivation of Eqs.~\eqref{eq:eom}-\eqref{eq:st} relied on the coupling between the deformations along the RBC membrane, and those normal to it (here depicted as $f_{l,m}$'s) through both membrane tension and the surrounding flow~\cite{vlahovska2011dynamics}. In the absence of flow, the linear elastic deformations induce forces only along the RBC membrane, while bending acts in the normal direction, leaving the membrane tension to balance both. The residual internal forces generated via the enucleation process balance both the forces along the RBC membrane and the normal forces, possibly inducing buckling via local compressive forces . This nontrivial coupling mechanism is inherently nonlinear~\cite{quilliet2006depressions}, and is currently missing from the framework --- possibly inducing the non-physical instabilities at low $\mcC$ values.

The analysis above demonstrates that an enucleated RBC can develop substantial negative membrane tension, which drives an instability. Membrane elasticity allows RBCs to hold memory both of their reference shape, here quantified by $\Delta_0$, and of their equilibrium shape, via elastic and bending forces that balance the membrane stresses. These two qualities allow RBCs to undergo, and sustain, an instability. A similar instability criterion was obtained for vesicles~\cite{turitsyn2008wrinkling}, understandably without the effects of the reference excess area $\Delta_0$, and without elasticity. In those cases, a sudden change in the external flow profile yielded negative membrane tension and a wrinkling instability, that was later on stabilized as the vesicle reoriented along the new flow direction~\cite{turitsyn2008wrinkling,kantsler2007vesicle}. When subjected to an elongation flow, such vesicles could develop pearling instabilities, under sufficiently high strain-rates~\cite{kantsler2008critical}. The situation here is different --- the instability persists, and allows for different modes to grow, resulting in non-trivial shapes, possibly performing oscillatory dynamics. In shear flow, in the absence of elastic forces, and in the absence of a mismatch $\Delta_0\neq\Delta$ sustaining negative tensions, this type of instability could not be attained.

\section{Post-instability excess area redistribution} \label{se:ss}

Once an instability is initiated, unstable modes grow at the expense of the other modes. This growth causes the membrane tension to increase until, eventually, the system reaches, or oscillates around, a steady state  --- the excess area must be conserved, so growth cannot persist indefinitely. The membrane tension saturates at $\sim\!\sigma_*$, as shown in Fig.~\ref{fig:fig3}(d)-(e), implying that all of the modes are now stable. The emerging post-instability shapes potentially correspond to the observed morphologies in experiments and simulations --- here we obtain analytical predictions for their spatial characteristics.

\begin{figure}[t]
\includegraphics[width=0.95\textwidth]{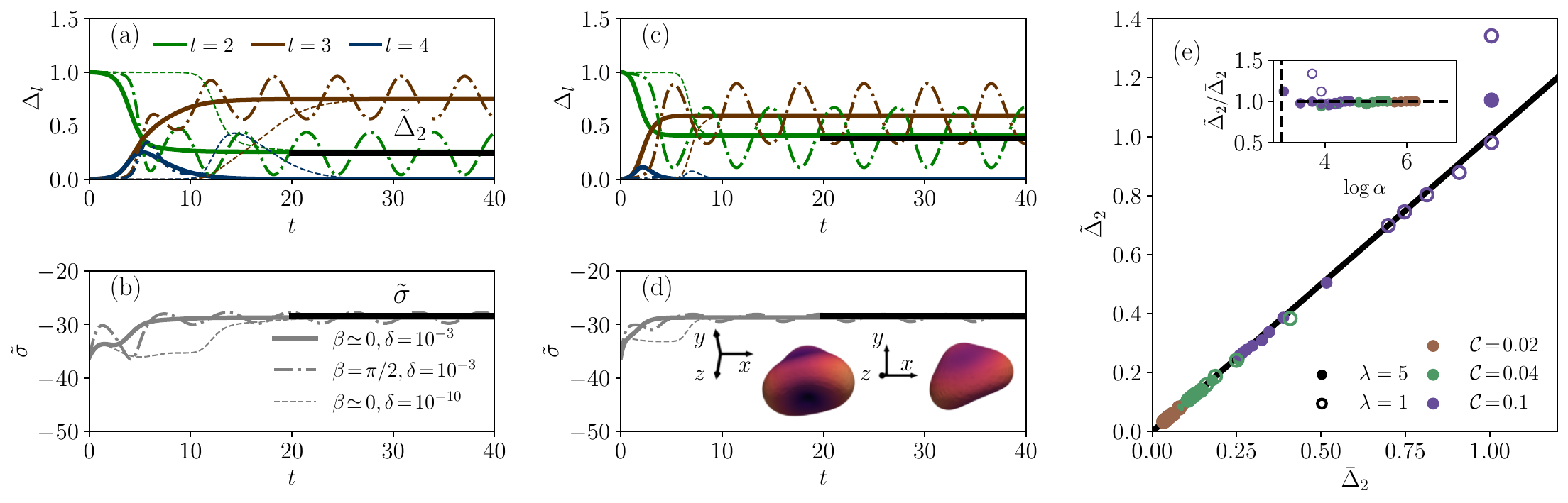}
\caption{Post-instability dynamics. (a) The excess areas $\Delta_l$ for $l\!=\!2\!-\!4$ for $\mcB\!=\!2$, $\mcC\!=\!0.04$, $\lambda\!=\!5$, $\Delta\!=\!1$, and $\Delta_0\!=\!0.01$. Solid, dashed, and dotted-dashed lines show the trajectories for $\Delta_{2,3,4}$ with different initial perturbation amplitudes of $\delta\!=\!10^{-3},10^{-10}$, and different orientations [see legend of (b)]. While the actual trajectories differ, the long-time state is similar. Solid black line shows the theoretical prediction for $\tilde{\Delta}_2$, in agreement with the steady-state $\bar{\Delta}_2$ attained by the simulations (shown in green). (b) The membrane tension for similar parameters as in (a), reemphasizing that while the dynamics differ, the steady-state values are approximately independent of $\delta$. (c)-(d) Same as (a)-(b) but for $\lambda\!=\!1$. Insets: the long-time configurations obtained from the OP orientation $\beta\!\simeq\!0$ (resulting in a stomatocyte) and the IP orientation $\beta\!=\!\pi/2$ (resulting in a trilobe) configurations. Note that the two configurations are plotted from a different angle (see their relative axes). Colors capture the distance from the origin. (e) A comparison between the theoretical $\tilde{\Delta}_2$ and the values obtained from the simulations $\bar{\Delta}_2$ for a range of $\mcC$ values, and $\mcB\!=\!1\!-\!10$ for both $\lambda\!=\!1$ and $\lambda\!=\!5$. An auxiliary diagonal line is added as a guide to the eye. Inset: the ratio $\tilde{\Delta}_2 / \bar{\Delta}_2$, as a function of $\log\alpha$, showing deviations as $\alpha$ decreases towards the instability boundary from above.}
\label{fig:fig6}
\end{figure}

Once a steady state is reached, the excess area is mainly distributed between the initial $l\!=\!2$ ellipsoidal modes, and $l\!=\!l_*$ modes. Denoting the long-time quantities with $\tilde{\bullet}$, we assume that $\Delta\simeq\tilde{\Delta}_2 + \tilde{\Delta}_{l_*}$, and that $\tilde{\sigma}(t)\simeq\sigma_*$. Then, we can approximate the values of $\tilde{\Delta}_2$ from Eq.\eqref{eq:st} as

\begin{equation}\label{eq:ss}
\tilde{\Delta}_{2}\simeq\frac{2\Gamma_{2}^{l_*}}{\Gamma_{2}^{2}\Gamma_{1}^{l_*}-\Gamma_{2}^{l_*}\Gamma_{1}^{2}+\alpha\Gamma_{2}^{2}\Gamma_{3}^{l_*}-\alpha\Gamma_{2}^{l_*}\Gamma_{3}^{2}}\mcB  \tilde{S}_{2m} \tilde{f}_{2m}^{*} \ .
\end{equation}
Importantly, this expression implies that the steady-state excess area distribution is independent of the initial perturbation amplitudes. We demonstrate this feature in Fig.~\ref{fig:fig6}(a)-(d), showing the $\Delta_l$ and $\sigma(t)$ dynamics for several cases of identical parameters, but with different initial perturbation amplitude $\delta$. Panels (a) and (c) show the $\Delta_l$ dynamics (for $l\!=\!2-4$) for different orientations $\beta$ and different initial perturbation amplitudes $\delta$, while panels (b) and (d) show the complementary membrane tension dynamics. Note that while the time to reach the post-instability regime depends on $\delta$, the final excess area distribution and membrane tension values (or their time-averaged equivalents) remain similar. 

The steady-state excess area Eq.~\eqref{eq:ss} requires the full solution, specifically knowing $\tilde{f}_{2,m}$ at long times (where, in principle, we need to know $\tilde{g}_{2,m}$ as well). To estimate $\tilde{f}_{2,m}$, we use substitutions for $\tilde{f}_{2,m}$ similar to those of Eq.~\eqref{eq:subs}, but now with $\tilde{\Delta}_{2}$ being the excess area within the $l\!=\!2$ mode, which is unknown in advance, $\tilde{r}$ as the new $r$ value, and $\tilde{\Psi}$ as the new $\Psi$ value attained post-instability. Explicitly, we take 
\begin{equation} \label{eq:subs_ss}
    \tilde{f}_{2,\pm2}\!=\!\sqrt{\frac{\tilde{\Delta}_{2}}{2 a_2}} \tilde{r} e^{\mp i \tilde{\Psi}}  \qquad \ , \qquad \tilde{f}_{2,0}\!=\!-\sqrt{\frac{\tilde{\Delta}_{2}}{a_{2}}\left(1-\tilde{r}^{2}\right)}  \ .
\end{equation}
Using the approximate membrane tension $\tilde{\sigma}(t)\!\simeq\!\sigma_*\!$, the equations of motion for $\tilde{r}$ and $\tilde{\Psi}$ in the reduced system after an instability are 
\begin{subequations}\label{eq:ssr}
    \begin{align}
        \tilde{r}' = & \tilde{\Lambda}^{-1}\sin\tilde{\Psi} + \left(\tilde{\Lambda}S\right)^{-1}r_0\cos\left(\tilde{\Phi}-\tilde{\Psi}\right) + \frac{2}{\sqrt{30\pi}}\tilde{\Lambda}^{-1}\left(\frac{6}{\sqrt{\mcB \mcC}}-\frac{1}{\mcC}-\frac{9}{\mcB}\right)\sqrt{\tilde{\Delta}_{2}} \tilde{r} \ , \\
        \tilde{\Psi}' = & -1+\tilde{\Lambda}^{-1}\frac{\cos\tilde{\Psi}}{\tilde{r}}+\left(\tilde{\Lambda} S\right)^{-1}\frac{r_{0}}{\tilde{r}}\sin(\tilde{\Phi}-\tilde{\Psi}) \ ,
    \end{align}
\end{subequations}
where we defined the post-instability $\tilde{\Lambda}\!\equiv\!\frac{\sqrt{\tilde{\Delta}_{2}} (23 \lambda +32)}{8 \sqrt{30 \pi }}$, depending on the unknown $\tilde{\Delta}_{2}$. While some of the terms are reminiscent of the unperturbed case of Eq.~\eqref{eq:red_eqs}, most of the terms are modified through $\tilde{\Lambda}$. There are additional terms in the $\tilde{r}$ equation that introduce an explicit $\mcB$-dependence, which is absent from the unperturbed system of Eq.~\eqref{eq:red_eqs}.

We now look for the steady state of $\tilde{r}$ and $\tilde{\Psi}$ for the OP case at the post-instability state described by Eq.~\eqref{eq:ssr}, and find that
\begin{equation}\label{eq:rss_postinst}
\tilde{r}_{ss}\!=\!\left[\left(\frac{16\left(9\mathcal{C}-6\sqrt{\mathcal{C}\mathcal{B}}+\mathcal{B}\right)}{\left(23\lambda+32\right)\mathcal{C}\mathcal{B}}\right)^2+1\right]^{-1/2} \tilde{\Lambda}^{-1} \ , \qquad \tilde{\Psi}_{ss}\!=\!\arctan\frac{16\left(9\mathcal{C}-6\sqrt{\mathcal{C}\mathcal{B}}+\mathcal{B}\right)}{\left(23\lambda+32\right)\mathcal{C}\mathcal{B}} \ .
\end{equation}
As $\tilde{f}_{2,\pm2}\!\sim\!\sqrt{\tilde{\Delta}_{2}}\tilde{r}\sim\sqrt{\tilde{\Delta}_{2}} \tilde{\Lambda}^{-1}$ and $\tilde{\Lambda}\sim\sqrt{\tilde{\Delta}_{2}}$, we do not need to know $\tilde{\Delta}_{2}$ to get the value of $\tilde{f}_{2,\pm2}$, however the value of $\tilde{f}_{2,0}$ still depends on $\sqrt{\tilde{\Delta}_{2}}$.  By substituting the initial values of $g_{2,m}$'s, and the new expressions for $\tilde{f}_{2,m}$ from Eq.~\eqref{eq:subs_ss}-\eqref{eq:rss_postinst} into Eq.~\eqref{eq:ss}, we obtain a quadratic equation for $\tilde{\Delta}_{2}$, which we solve numerically, for a range of $\mcB$ and $\mcC$ parameters. 

The predicted $\tilde{\Delta}_2$ values are shown in thick black lines in Fig.~\ref{fig:fig6}(a),(c), and $\tilde{\sigma}(t)\!=\!\sigma_*$ are shown as thick black lines in the corresponding panels (b) and (d). We show the systematic agreement between our theoretical estimate $\tilde{\Delta}_{2}$ and the numerically-obtained values denoted here as $\bar{\Delta}_2$ in Fig.~\ref{fig:fig6}(e). Deviations are observed as $\alpha$ approaches the stability boundary from above, as shown in the inset of Fig.~\ref{fig:fig6}(e).

Finally, as the effective instability occurs for $\Delta_l$ and not for individual $f_{l,m}$ component (see Appendix ~\ref{app:theory}), the long-time structures inherit their symmetry from their initial state and their orientation with respect to the flow. We present two examples of emerging shapes in the insets of Fig.~\ref{fig:fig6}(d) --- an initially OP RBC evolves into a stomatocyte (left inset), while an IP RBC evolves into a multi-lobed shape (right inset). This emerges due to our assumption about the conservation of axial symmetry during the enucleation process. This is one possible explanation for the emergence of different RBC morphologies under the same external flow profiles, reported in~\cite{lanotte2016red}.

\section{Discussion} \label{se:discussion}
Inspired by recent observations of large deformations in RBCs under shear flow~\cite{lanotte2016red,mauer2018flow,levant2016intermediate}, we extended a perturbative framework~\cite{vlahovska2011dynamics} by first including different excess areas for the stress-free and current shapes, $\Delta_0$ and $\Delta$, respectively. We first observed that in the ellipsoidal approximation, where only $l\!=\!2$ modes are considered, the emerging dynamics is reminiscent of the case with $\Delta\!=\!\Delta_0$, with a modification to the governing parameter $S$. In the present framework, $S$ encapsulates not only the elastic properties of the RBC membrane through $\mcC$, but also a memory of its stress-free shape through $\Delta_0$.

Then, we incorporated additional spherical harmonics as perturbations, and examined the emerging dynamics. We derived an instability criterion, predicted the most prone spherical harmonics for growth, and approximated the stability boundaries as the RBC's mechanical properties are changed. In the present framework, the membrane membrane tension is the only nonlinear coupling mechanism driving the instability, consistent with the discussion from Abbasi \emph{et. al}~\cite{abbasi2022dynamics}. However, our stability analysis allows the identification of the role of membrane bending resistance and elasticity: we demonstrated the effects of the bending number $\mcB$ on the RBC's stability (absent from the ellipsoidal equations of motion), and showed that $\alpha\!=\!\mcB/\mcC$, the ratio of bending and elastic forces, controls the excited harmonics: modes of higher $l$ values are excited as $\alpha$ increases. We have also demonstrated that changing the viscosity ratio $\lambda$ or the excess area $\Delta$ can change the stability boundaries, mainly through the dynamics of the membrane tension. We also probed the stability in $\log \mcC - \lambda$ space, and obtained results reminiscent of those obtained from full numerical solutions~\cite{mauer2018flow}.

Finally, the simplicity of the model allowed us also to understand the post-instability long-time behaviors. An instability induces a redistribution of the excess areas, until the membrane tension $\sigma(t)$ stabilizes at $\sigma_*$. Once this state is reached, we assumed only the $l\!=\!2$ and $l_*$ modes contribute to the excess area, and derived the effective reduced equations of motion after an instability. Using the expected steady-state solutions, we found an estimated excess area for the $l\!=\!2$ subspace, which we confirmed numerically.  As a result of the initial axi-symmetric perturbation, different three-dimensional shapes emerged under the same flow conditions, as also observed in experiments and simulations~\cite{lanotte2016red}. Combining the initial conditions with the value of $\alpha$ and our stability predictions allowed us to understand the long-time dynamics and to demonstrate that $\alpha$ controls the emerging shape (e.g., the spherical harmonics that become prominent) while the initial conditions control the emerging dynamics (e.g., whether the long-time trajectories reach a steady state or exhibit oscillations).

Experimentally, RBCs are known to adopt morphologies different from their typical discocyte shape, specifically when subjected to intermediate and large shear rates~\cite{lanotte2016red,levant2016intermediate}. The theoretical framework proposed above constitutes a first step in understanding such type of instabilities. For example, experimentally, in steady shear flow, trilobes were observed for
$\dot{\gamma}\sim 1000\ s^{-1}$ and $\eta^{\text{ex}}\simeq1\ \text{cP}$.
Taking $R\simeq3\ \mu m$, $\mu\!\simeq\! 2.5\ \mu \text{Pa}\cdot\text{m}$ and $\kappa\!\simeq\!2.0 \times 10^{-19} \ \text{J}$ yields $\mcB\!\sim\!10^2$ and $\mcC\!\sim\!1$ ($\log\!\mcB\!\approx\! 4.5$ and $\log\mcC\!\approx\!0$), which are in the unstable region at $\lambda\!=\!5$ (Fig.~\ref{fig:fig4}b). 
However, sustained stomatocyte and trilobe morphologies appeared in the experiments above a critical shear rate~\cite{lanotte2016red} --- a feature not captured by our theoretical framework, indicating some elements are still missing. The proposed theoretical framework suggests additional dependencies on $\lambda$, $\alpha$ and $\Delta$, which were not probed experimentally yet. 

To further understand the instability mechanisms, and to test the theory above, additional experiments of diluted RBCs or single RBCs are required. Modifying the shear rate $\dot{\gamma}$ and the external viscosity $\eta^{\text{ex}}$ should allow probing the significance of the viscosity ratio $\lambda$ to the instability (see Fig.~\ref{fig:fig5} and the related discussion) --- the instabilities are expected to disappear as $\lambda$ decreases. Modification to the RBC membrane (e.g., using membrane-hardening techniques~\cite{lanotte2016red}) should allow probing the importance of elastic forces --- though as mentioned, the instabilities for low $\mcC$ and low $\mcB$ values are unphysical. Probing RBCs with different excess areas (e.g., obtained using osmotic pressure effects), would help elucidate the role of the locked internal stresses and elasticity, where generally the more deflated is the RBC, the more susceptible it becomes to instabilities.  Fine control over the initial orientation would also allow clarifying the role of the initial orientation with respect to the external flow. Probing the RBCs surface and whether the deformed shapes are sustained or intermittent could shed light on the underlying differences between vesicles and RBCs.

While the theoretical approach above allowed a thorough analysis of mechanical instabilities, this approach also lacks some aspects that probably play an important role in real RBC dynamics. First, the entire approach is based on perturbative expansion from a spherical shape. Clearly, when an RBC undergoes enucleation, the emerging discocyte is very different from a sphere --- this geometric difference may be crucial in the emerging dynamics. Additionally, we have taken moderate $\Delta$ values, which may be insufficiently small, causing the perturbative assumption to break down. In realistic RBCs $\Delta\!\simeq\!4$~\cite{vlahovska2011dynamics,evans1972improved,levant2016intermediate}, and it remains questionable whether the perturbative theory respects such values. 

Second, as the theory is approximately linear, by construction it lacks some possibly important non-linear contributions. Currently, the membrane tension term is the only nonlinear coupling in the emerging equations. Yet, it is clear that the interactions between RBCs and their surrounding flow is an inherently non-linear problem --- the flow deforms the RBC membrane that modifies the flow again. Additionally, the RBC membrane itself has a non-linear response (due to its initial biconcave geometry, and its non-linear constitutive response. These effects may be important for the instability onset --- for example, nonlinear elastic contributions could excite higher harmonics and induce an instability in regions rendered as stable by the above theory. In the long-time limit, coarsening effects could smooth high $l$ perturbations towards lower $l$ values possibly altering the long-time sustained shape.

Finally, the theoretical description above assumes that the stress-free shape does not reorient during its dynamics, and maintains a constant orientation with respect to the flow plane. This limits the current theoretical framework to a fixed angle with respect to the flow. Statistically, RBCs orientation with respect to the external flow is not well controlled, and could induce competition between rotation towards/from the flow plane and an instability growth. For example, in~\cite{levant2016intermediate} the rate of rotation clearly played a crucial role in inducing instabilities, but those were not sustained (hence termed ``intermittent''). Capturing the rotation of the stress-free structure with respect to the ambient flow field could induce new dynamical behaviors and enrich the stability analysis.

Overall, the simplicity of the perturbative framework reported here allowed us to gain many insights into the instability onset and dynamics of RBCs under shear flow. These observations serve as first steps in understanding RBC dynamic instabilities, and could motivate additional numerical and experimental investigations into the instability and post-instability dynamics of RBCs in flow. Such future directions would help understand the emergence of various RBC shapes, elucidate the roles played by the RBC membrane mechanics on the emerging blood rheology, and potentially provide insights into biologically relevant problems, such as the identification of membrane anomalies from the observation of RBC dynamics.

\section{Acknowledgments}

A.M. acknowledges support from the \href{https://doi.org/10.37717/2021-3362}{James S. McDonnell Foundation Postdoctoral Fellowship Award in Complex Systems}. The authors gratefully acknowledge the computing resources provided by GENCI (under grant allocations A0160307194 and A0180307194) and by the ISDM computing center (Montpellier, France). The authors also thank Dr. Manouk Abkarian (CNRS Montpellier), and the two anonymous reviewers, for helpful discussion and comments. 

\appendix

\section{Spherical harmonics and Wigner D matrix}
\label{app:Ylmrot}
We parameterize the RBC surface using an expansion in the spherical harmonics $Y_{l,m}$'s, as defined in~\cite{vlahovska2011dynamics}
\begin{equation}
    Y_{l,m}(\theta,\varphi) = \left[\frac{2l+1}{4\pi}\frac{(l-m)!}{(l+m)!}\right]^{1/2} (-1)^m P_l^m\left(\cos\theta\right)e^{im\varphi} \ ,
\end{equation}
where $P_l^m\left(\cos\theta\right)$ are the Legendre polynomials.

As mentioned in the main text, we first initiate an axisymmetric RBC structure, parameterized by $Y_{l0}$'s only. This corresponds to a structure oriented out of plane (OP). To rotate this structure, we need to apply the rotation operator $\mcR\left(\beta_1,\beta_2,\beta_3\right)$ defined here as
\begin{equation}
    \mcR\left(\beta_1,\beta_2,\beta_3\right)\!=\!e^{-i \beta_1 J_z} e^{-i \beta_2 J_y} e^{-i \beta_3 J_z} \ ,
\end{equation}
where the $J$'s are the generators of the $\text{SO}(3)$ Lie algebra, and $\beta$'s are the Euler angles (using the $z\!-\!y\!-\!z$ convention). The Wigner D-matrix is the rotation $\mcR$ projected onto the specific subspace of interest, defined here as
\begin{equation}
D^{(l)}_{m',m}\left(\beta_1,\beta_2,\beta_3\right)\!\equiv\!\langle l,m'|\mcR\left(\beta_1,\beta_2,\beta_3\right)|l,m\rangle\ .
\end{equation}
Here we use rotations in the $y$ direction only, simplifying the dependence of $\mcR$ and $D$ to depend on $\beta_2\!\equiv\!\beta$ (sometimes referred to as Wigner's d-matrix).

For example, the entries of $D^{(2)}_{m,0}(\beta)$ are obtained~\cite{Mathematica} as
\begin{equation*}
    D^{(2)}_{0,0}\left(\beta\right) = \frac{1}{4}\left(3\cos 2\beta+1\right)\ , \qquad D^{(2)}_{\pm1,0}\left(\beta\right) = \pm\sqrt{\frac{3}{2}}\cos\beta\sin\beta \ , \qquad D^{(2)}_{\pm2,0}\left(\beta\right) = \frac{1}{2}\sqrt{\frac{3}{2}}\sin^2\beta\ .
\end{equation*}

\section{The perturbative framework}
\label{app:theory}

The theoretical framework considered in the main text results from a systematic perturbation expansion of the full equations of motion~\cite{vlahovska2011dynamics,vlahovska2007dynamics,olla2000behavior}. We consider the internal and external fluids as incompressible Newtonian fluids with negligible Reynolds number (mainly due to the spatial scale of the RBC). The external fluid is driven at infinity as $\bm{v}^{\infty}\!=\!\dot{\gamma} y \hat{x}$. At the RBC boundary, the fluid velocities identify with the membrane velocity, and the emerging stress difference between the external and internal fluids is balanced by the membrane stresses.

The full details and corresponding equations of motion are provided in~\cite{vlahovska2011dynamics}. The perturbative expansion eventually yields the equations of motion for $f_{l,m}$'s as
\begin{equation} \label{eq:flms}
\partial_{t}f_{l,m}\!=\!\left[i\omega m+\Upsilon _l(t)\right]f_{l,m}+S_{l,m}(t) \ ,
\end{equation}
where
\begin{subequations} 
\begin{align}
    \Upsilon_l(t) = & \mcB^{-1}\left[\Gamma_1^l + \sigma(t) \Gamma_2^l\right] + \mcC^{-1} \Gamma_3^l \ ,\\ 
    S_{l,m}(t) = & C_{l,m} - \mcC^{-1} \Gamma_3^l g_{l,m} \ ,
\end{align}
\end{subequations}
and
\begin{subequations} \label{eq:gammas}
\begin{align}
    d(\lambda,l) = & (2l^3+3l^2+4) + \lambda(2l^3 + 3l^2 - 5) \ , \\
    \Gamma_1^l = & -\frac{\left(l+2\right)\left(l+1\right)^2l^2\left(l-1\right)}{d(\lambda,l)} \ , \\ 
    \Gamma_2^l = & -\frac{\left(l+2\right)\left(l+1\right)l\left(l-1\right)}{d(\lambda,l)}\ , \\
    \Gamma_3^l = & -\frac{4\left(l+2\right)\left(l-1\right)}{d(\lambda,l)} \ .
\end{align}
\end{subequations}
The external driving contributes 

\begin{equation}
    C_{l,m} = d(\lambda,l)^{-1}\left[c_{l,m,0}^\infty\sqrt{l(l+1)}(2l+1) + c_{l,m,2}^\infty(4l^3 + 6l^2 - 4l -3) \right] \ .
\end{equation}

In the case of shear, the only components that contribute to $c_{l,m,0}^\infty$ and $c_{l,m,2}^\infty$ are $c_{2,\pm2,0}^\infty\!=\!\mp i \sqrt{\pi/5}$ and $c_{2,\pm2,2}^\infty\!=\!\mp i \sqrt{2\pi/15}$. 

Finally, the membrane tension $\Sigma(t)$ is also expressed using spherical harmonics as
\begin{equation}
\Sigma(\theta, \phi, t) = \sigma(t) + \sum \sigma_{l,m}(t) Y_{l,m}\left(\theta,\varphi\right) \ .    
\end{equation}
Solving for the tangential stress balance specifies the values of $\sigma_{l,m}$~\cite{vlahovska2011dynamics}, which were already substituted into the normal equations of motion, Eq.~\eqref{eq:flms}. The value of the spatially-homogeneous part of the membrane tension $\sigma(t)$ is
obtained by requiring a constant excess surface area $\partial_t\Delta \!=\! 0$, which yields Eq.~\eqref{eq:st}.

The expression for $\Upsilon_l$, Eq.~\eqref{eq:eom_sup_ups}, clearly demonstrates that the growth rate depends on $l$ via the $\Gamma$ functions, but not on $m$. This implies that within a certain $l$ subspace, perturbations grow proportionally to their initial value. Under shear, and assuming that the stress-free shape has only ellipsoidal components, $S_{l,m}$ vanishes, and the dynamics reduces to $\partial_t \Delta_l \!=\!2\Upsilon_l \Delta_l$. This implies that for $l\!\ge\!3$ one can examine the instability through $\Delta_l$.

\begin{figure}[h!]
\includegraphics[width=0.9\textwidth]{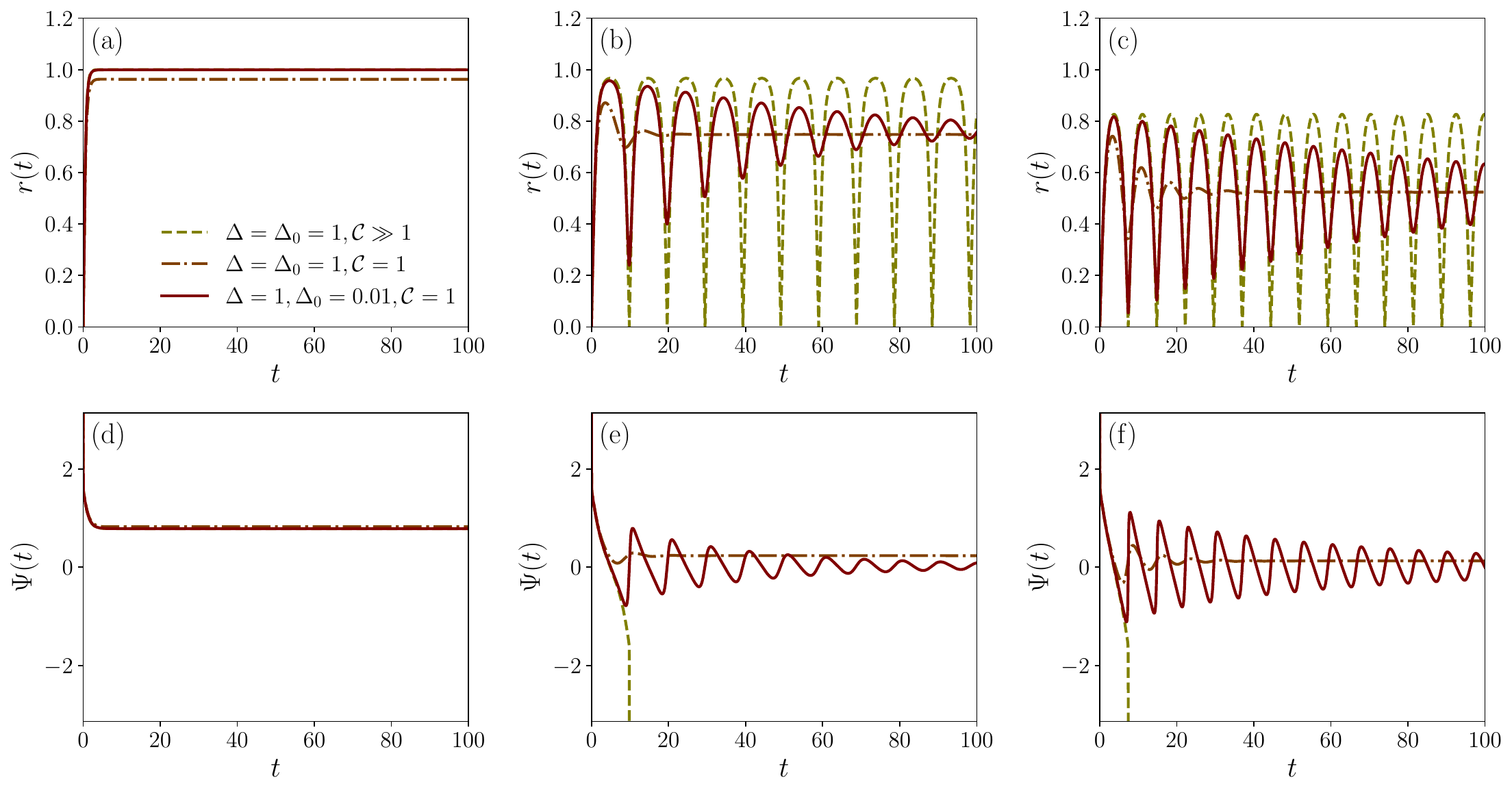}
\caption{Unperturbed dynamics of out-of-plane vesicle-like RBC (in dashed green lines), unstressed RBC (in dotted-dashed light blue lines), and enucleated RBC (in blue lines) under different viscosity ratios $\lambda$. Top panels show the amplitude dynamics $r(t)$, and the bottom panels show the angular dynamics $\Psi(t)$. Panels (a) and (d) correspond to $\lambda\!=\!1$, (b) and (e) to $\lambda\!=\!3$, and (c) and (f) to $\lambda\!=\!5$. The dynamics are discussed above, however in general the enucleated RBC exhibits dynamics in-between the vesicle-like RBC and the unstressed RBC.}
\label{fig:app3}
\end{figure}

\section{Survey of unperturbed dynamics}\label{app:vesicle}

The reduced system detailed in Sec.~\ref{se:upd}, with $\Delta\!=\!\Delta_0$ was thoroughly developed and studied in~\cite{vlahovska2011dynamics}. Here, we do not intend to perform a thorough investigation of the underlying available dynamics of our modified equations of motion Eq.~\eqref{eq:red_eqs} (such analyses were performed in~\cite{vlahovska2011dynamics}), but rather to showcase various dynamics emerging from the current formalism, and compare and contrast those with the corresponding vesicle dynamics.

We show in Fig.~\ref{fig:app3} the emerging $r$ and $\Psi$  dynamics, in the top and bottom panels respectively  for three values of $\lambda$, when the RBC is oriented out-of-plane. Panels (a),(d) show the dynamics for $\lambda\!=\!1$, (b),(e) for $\lambda\!=\!3$, and (c),(f) for $\lambda\!=\!5$. In each panel, we show the case of large $\mcC\!=\!10^6$ value, corresponding to large $S$ value, which is the vesicle limit, low $C\!=\!1$ value obtained with $\Delta\!=\!\Delta_0$, and finally the case of low $S$ value with $\Delta\neq\Delta_0$.

\begin{figure}[h!]
\includegraphics[width=0.9\textwidth]{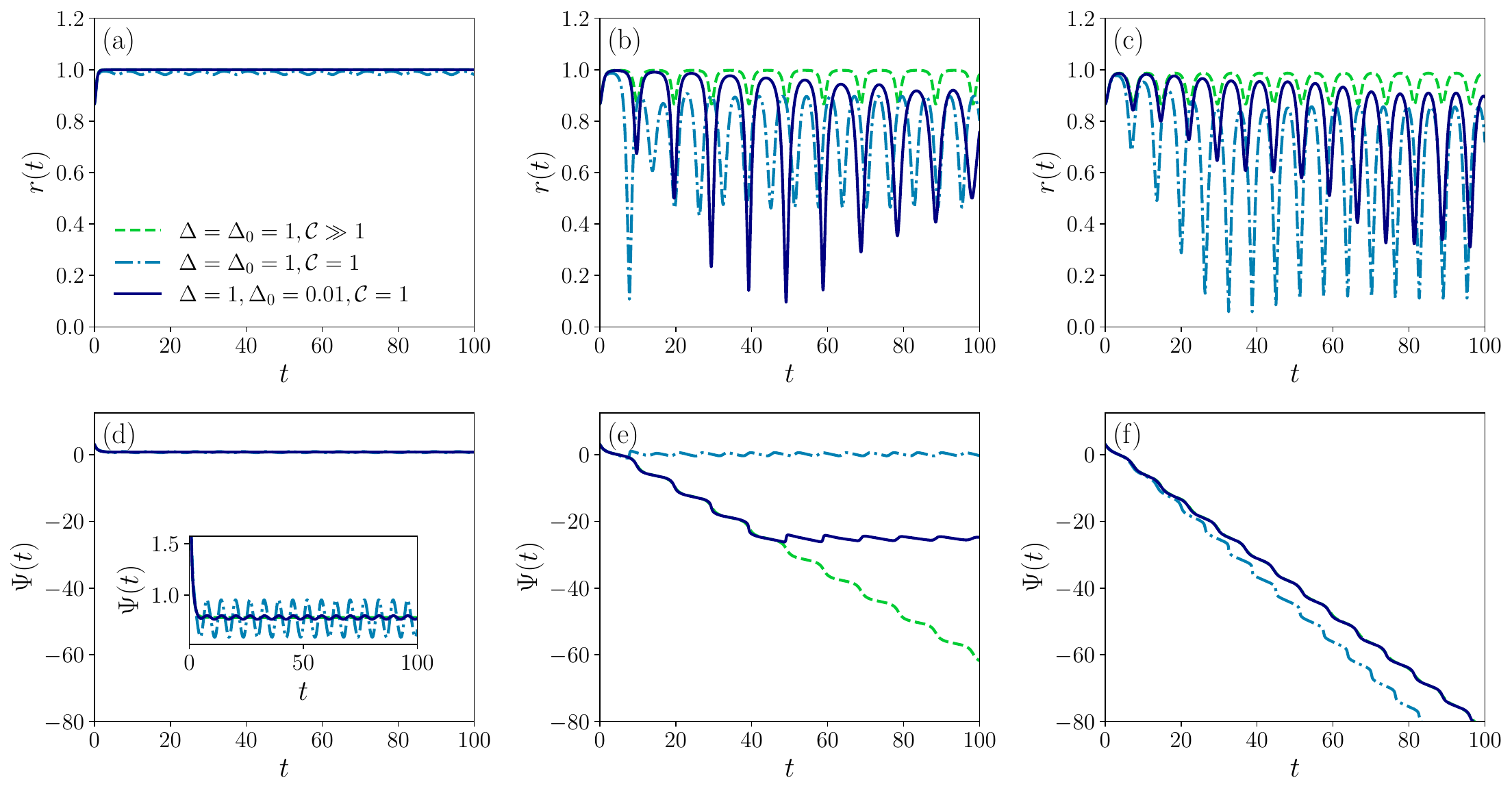}
\caption{Unperturbed dynamics of in-plane vesicle-like RBC (in dashed green lines), unstressed RBC (in dotted-dashed brown lines), and enucleated RBC (in red lines) under different viscosity ratios $\lambda$. Top panels show the amplitude dynamics $r(t)$, and the bottom panels show the angular dynamics $\Psi(t)$. Panels (a) and (d) correspond to $\lambda\!=\!1$, (b) and (e) to $\lambda\!=\!3$, and (c) and (f) to $\lambda\!=\!5$. Inset of panel d: a zoom in showing small angle oscillations for the unstressed RBC. The dynamics are discussed above, however in general the enucleated RBC exhibits dynamics in-between the vesicle-like RBC and the unstressed RBC.}
\label{fig:app4}
\end{figure}

The dynamics for $\lambda\!=\!1$ are almost indistinguishable --- all cases converge to tank-treading along $\Psi\!=\!\pi/2$, corresponding to $\psi\!=\!\pi/4$. For higher $\lambda$ values, the vesicle-like RBC tumbles and deforms the most [see panels (b)-(c)], while the unstressed RBC converges to a steady-state quickly [at $t\simeq20$ in (b), and $t\simeq40$ in (c)]. Interestingly, the enucleated RBC exhibits damped oscillations, but the damping occurs on a longer time-scale, showing an intermediate dynamics between the vesicle case and the $\Delta\!=\!\Delta_0$ case.

We repeat a similar examination for the in-plane case in Fig.~\ref{fig:app4}. Again the dynamics for $\lambda\!=\!1$ are similar across the three different cases, with the exception of small angle oscillations for the unstressed RBC. This is the signature of swinging~\cite{abkarian2007swinging} --- the elasticity of the membrane modulating its circulation and thus the particle inclination. These oscillations are reduced when having $\Delta_0\!<\!\Delta$, consistently with previous studies \cite{dupire2015simple,mendez2018plane}. For $\lambda\!=\!3$, both the vesicle-like RBC and the unstressed RBC perform amplitude oscillations in $r(t)$, where again the internally-stressed RBC exhibits a dynamical behavior in-between the vesicle-like RBC and the unstressed RBC. It exhibits decaying amplitude oscillations, where the decay occurs on time-scales longer than those of the unstressed RBC. The vesicle-like RBC tumbles, the unstressed RBC performs swinging, and the stressed RBC starts tumbling, and transitions into a swinging (the averaged angle is non-zero). When $\lambda$ is further increased, all three RBCs perform deformed tumbling.

\section{Steady-state solutions and derived instability criteria}
\label{app:reducedsteadystates}

The reduced system of Eq.~\eqref{eq:red_eqs} admits several steady-state solutions. Here we focus on the leading-order steady-state solution $r_{ss}^{-}$,$\Psi_{ss}^{-}$ when $\Lambda\!<\!1$, and $r_{ss}^{+}$, $\Psi_{ss}^{+}$ for $\Lambda\!>\!1$, provided in Eq.~\eqref{eq:ss_red} (detailed in~\cite{vlahovska2011dynamics}). To obtain the instability criteria for each of these cases, we use the above solutions in $\sigma_R(t)$ and perform time averaging $\langle \sigma_R^{\bullet} \rangle\!\equiv\!\frac{1}{2\pi}\int_0^{2\pi}\sigma_R^{\bullet}(t')dt'$, allowing us to eliminate the $\Phi$ dependence in Eq.~\eqref{eq:st_red}. This yields analytical expressions for the membrane tension for the weak elastic scenarios, as
\begin{subequations} \label{eq:ss_st_red}
    \begin{alignat}{2}
    \langle \sigma_R^{-} \rangle & = -6+\frac{\sqrt{30\pi}}{3}\sqrt{\frac{1-\Lambda^{2}}{\Delta}}\mathcal{B}-\frac{2\mathcal{B}}{3\mathcal{C}}\quad , &\quad \text{for }\Lambda<1 , \\
\langle \sigma_R^{+} \rangle&= -6+\frac{2\mathcal{B}}{3\mathcal{C}}\left(\sqrt{\frac{\Delta_{0}}{\Delta}\frac{\left(1-\Lambda^{2}\right)\left(1-r_0^{2}\right)}{\Lambda^{2}}}-1\right) \quad , & \quad \text{for }\Lambda>1\ .
    \end{alignat}
\end{subequations}
Note that $\langle \sigma_R^{-} \rangle$ depends on $\mcB$, and not only on the ratio $\mcB/\mcC$, hinting that the derived criterion may have a non-trivial dependence on $\mcB$. Additionally, note that $\langle \sigma_R^{+} \rangle$ depends explicitly on $r_0$, implying the instability criterion depends on the initial orientation. Finally, we demand that $\langle \sigma_R^{\bullet} \rangle\!\le\!\sigma_*$ for the expressions in Eq.~\eqref{eq:ss_st_red}, allowing us to obtain the critical $\mcC_*$ values for instability shown in Eq.~\eqref{eq:c_crit}. Note that to obtain a similar criterion for $\lambda_*(\mcC)$ for fixed $\alpha$ values, the $\Lambda\!<\!1$ case yields a non-trivial dependence due to an explicit dependence on $\mcB$, while the $\Lambda\!>\!1$ case depends on $\mcC$ only through $\alpha$, yielding a $\mcC$-independent $\lambda_*$.

\section{In-plane stability}
\label{app:in_and_out}
\begin{figure}[t]
\includegraphics[width=0.97\textwidth]{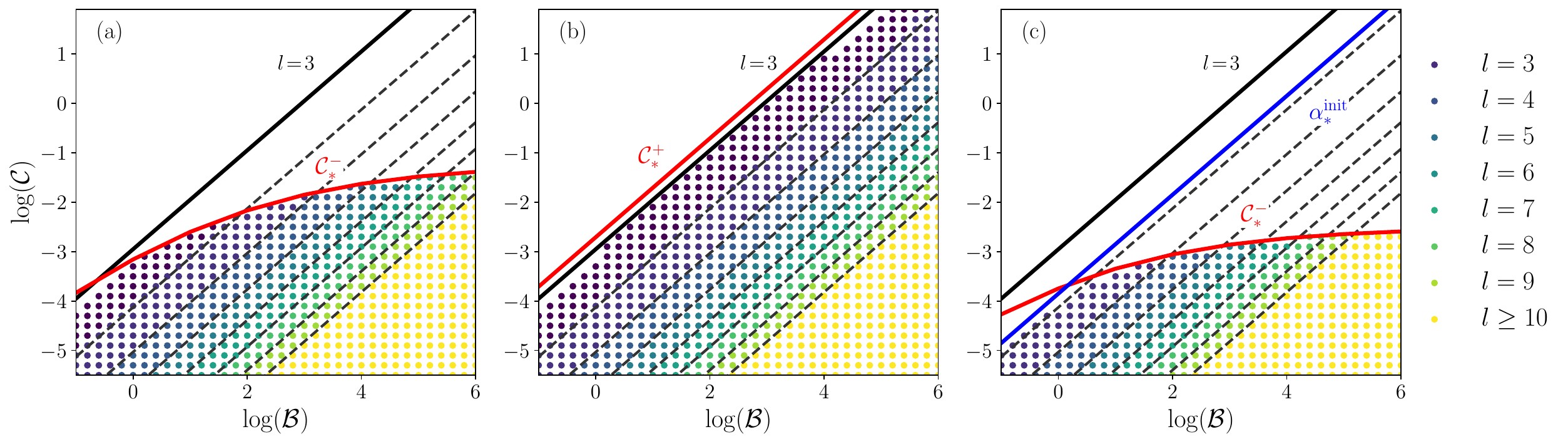}
\caption{IP stability phase diagrams similar to those of Fig.~\ref{fig:fig4}, for different $\lambda$ values [panels (a)-(b)], and different $\Delta$ values [panels (b)-(c)]. Log-log phase diagrams for (a) $\lambda\!=\!1$, $\Delta\!=\!1$, (b) $\lambda\!=\!5$, $\Delta\!=\!1$ and (c) $\lambda\!=\!5$, $\Delta\!=\!0.1$ in the IP orientation ($\Delta_0\!=\!0.01$ for all cases). Each phase diagram was obtained by sampling $\left(\mcB,\mcC\right)$ pairs spanning $\mcB\!\in\!\left[\exp(-1),\exp(6)\right]$, and $\mcC\!\in\!\left[\exp(-5.5), \exp(2)\right]$, and simulating each perturbed system for a duration of $T\!=\!500$. Colors indicate the value of $l_\Delta$, indicating the $l$ value for which the excess area increased the most, as obtained from the numerical simulations (see legend). Solid diagonal line show the theoretical predictions for $l_*\!=\!2.5$ (where we expect $l\!=\!3$ to be excited), and the additional diagonal dashed lines show the theoretically-predicted transition from excited $l$ value to the next one. Red curves show the value of $\mcC_*$ obtained from the steady-state calculation (see Appendix~\ref{app:reducedsteadystates}). Finally, blue diagonal shows $\alpha_c^{\text{init}}$ effecting the instability boundary at low $\mcB$ and $\mcC$ values [$\alpha_c^{\text{init}}$ is not shown in panels (a) and (b) as it roughly overlaps with the $l\!=\!3$ critical line, as mentioned above]. The emerging picture is similar to Fig.~\ref{fig:fig4}, where RBCs of high $\mcC$'s are destabilized by high viscosity contrast $\lambda$ and large excess area $\Delta$.}
\label{fig:app1}
\end{figure}

We probed the $\mcB$-$\mcC$ phase spaces for the IP cases, in a similar fashion to the analysis presented in Sec.~\ref{se:inst}. Here, we had to define an average observable as the IP cases yield oscillatory dynamics [e.g., see Fig.~\ref{fig:fig3}(c)]. Probing only the last entry of $\Delta_l$ in a given trajectory is prone to error, as it depends critically on where this last point occurred relative to the oscillation cycle. To circumvent this issue, we defined $\langle \Delta_l\rangle$ as an average over the last $10$ sampled entries obtained from the numerical simulations (corresponding to $t\in 490-500$). The emerging phase diagrams, complementary to those presented in Fig.~\ref{fig:fig4}, are plotted in Fig.~\ref{fig:app1}, showing similar results.

\begin{figure}[h]
\includegraphics[width=0.7\textwidth]{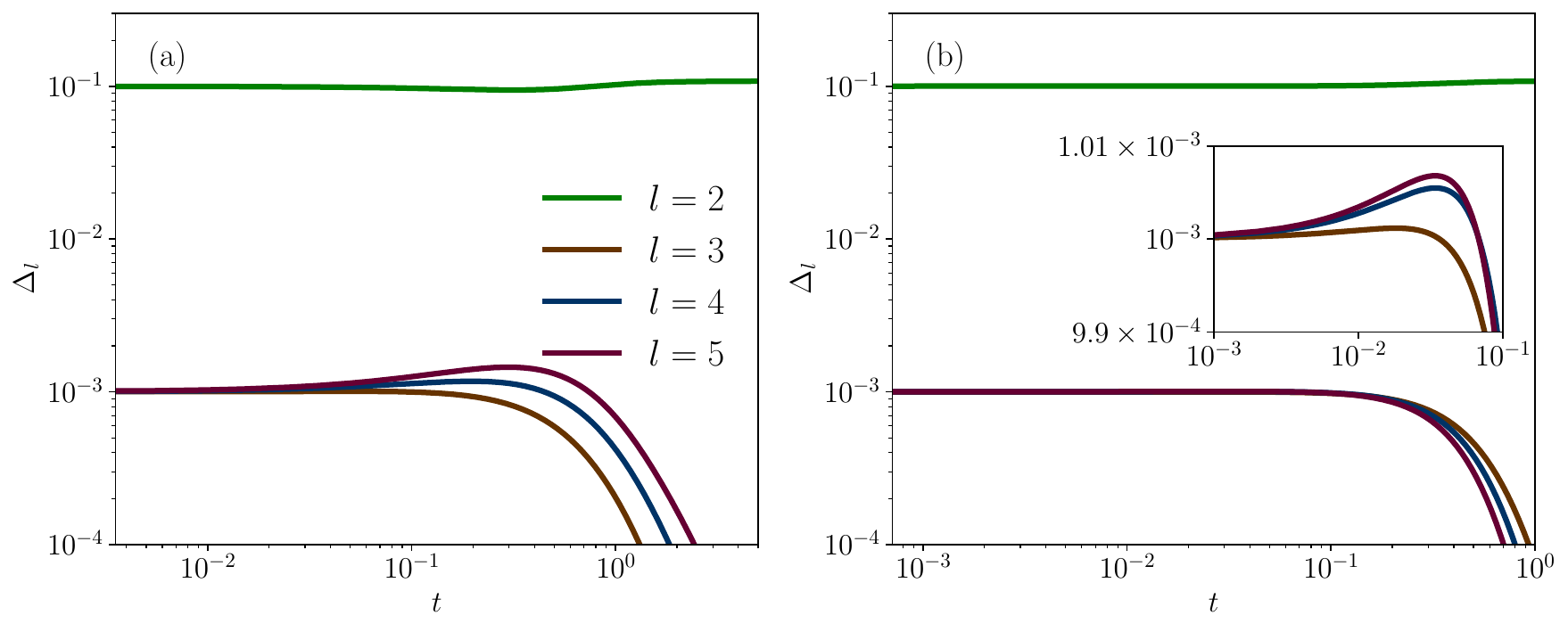}
\caption{Suppressed initial instability. (a) Initial dynamics for $\Delta\!=\!0.1$, $\Delta_0\!=\!0.01$, $\lambda\!=\!5$, $\mcB\!=\!\exp(4)$, $\mcC\!=\!\exp(-2.7)$ showing an initial growth suppressed by the underlying reduced dynamics. (b) Similar plot for $\mcB\!=\!\exp(4)$, $\mcC\!=\!\exp(-1)$ (all other parameters kept the same).}
\label{fig:app2}
\end{figure}

An interesting behavior occurs for $\Delta\!=\!0.1$, $\lambda\!=\!1$, corresponding to Fig.\ref{fig:fig4}(c) and Fig.~\ref{fig:app1}(c). There are regions that are initially unstable but become stable at the steady state of the reduced system (occurring for high $\mcB$ and $\mcC$ values, in the region above $\mcC_*^{-}$ but below $\alpha_{*}^{\text{init}}$). Then, there are also cases that are initially stable, but are destabilized in the steady state (occurring for low $\mcB$ and $\mcC$ values, in the region above $\alpha_{*}^{\text{init}}$ but below $\mcC_*^{-}$). We provide two exemplary trajectories from the first scenario in Fig.~\ref{fig:app2} --- initially unstable trajectories are stabilized by the approach to the steady-state solution of the reduced system. The second scenario is much harder to observe as the initial stability causes the perturbations to decay rapidly. 

\end{document}